\documentclass[twocolumn,amssymb,superscriptaddress]{revtex4-2}
\usepackage{amsmath}
\usepackage{amssymb}
\usepackage{amsthm}
\usepackage{amsfonts}
\usepackage[version=4]{mhchem}
\usepackage{listings}
\usepackage{enumerate}
\usepackage{latexsym}
\usepackage{psfrag}
\usepackage{bm} 
\usepackage[all]{xy}
\usepackage{graphicx}
\usepackage{subfigure}
\usepackage[pdftex,colorlinks]{hyperref}
\usepackage{color}
\usepackage{mathtools}
\usepackage{times}
\usepackage{array}
\usepackage{placeins}

\begin{document}

\title{ Dynamic fingerprint of fractionalized excitations in single-crystalline \ce{Cu3Zn(OH)6FBr}}
\author{Ying Fu}
\thanks{These two authors contributed equally.}
\affiliation{Shenzhen Institute for Quantum Science and Engineering, and Department of Physics, Southern University of Science and Technology, Shenzhen 518055, China}
\author{Miao-Ling Lin}
\thanks{These two authors contributed equally.}
\affiliation{State Key Laboratory of Superlattices and Microstructures, Institute of Semiconductors, Chinese Academy of Sciences, Beijing 100083, China}
\affiliation{Center of Materials Science and Optoelectronics Engineering \& CAS Center of Excellence in Topological Quantum Computation, University of Chinese Academy of Sciences, 100049 Beijing, China}
\author{Le Wang}
\author{Qiye Liu}
\author{Lianglong Huang}
\author{Wenrui Jiang}
\author{Zhanyang Hao}

\author{Cai Liu}
\affiliation{Shenzhen Institute for Quantum Science and Engineering, and Department of Physics, Southern University of Science and Technology, Shenzhen 518055, China}
\author{Hu Zhang}
\author{Xingqiang Shi}
\affiliation{College of Physics Science and Technology, Hebei University,
Baoding 071002, China}
\author{Jun Zhang}

\affiliation{State Key Laboratory of Superlattices and Microstructures, Institute of Semiconductors, Chinese Academy of Sciences, Beijing 100083, China}
\affiliation{Center of Materials Science and Optoelectronics Engineering \& CAS Center of Excellence in Topological Quantum Computation, University of Chinese Academy of Sciences, 100049 Beijing, China}
\author{Junfeng Dai}
\author{Dapeng Yu}
\affiliation{Shenzhen Institute for Quantum Science and Engineering, and
  Department of Physics, Southern University of Science and Technology, Shenzhen
  518055, China}
\author{Fei Ye}
\affiliation{Shenzhen Institute for Quantum Science and Engineering, and
  Department of Physics, Southern University of Science and Technology, Shenzhen
  518055, China}
 \affiliation{Shenzhen Key Laboratory of Advanced Quantum Functional Materials
   and Devices, Southern University of Science and Technology, Shenzhen 518055, China}

\author{Patrick A. Lee}
\affiliation{Department of Physics, Massachusetts Institute of Technology,
  Cambridge, Massachusetts 02139, USA}
\author{Ping-Heng Tan}
\email{phtan@semi.ac.cn}
\affiliation{State Key Laboratory of Superlattices and Microstructures, Institute of Semiconductors, Chinese Academy of Sciences, Beijing 100083, China}
\affiliation{Center of Materials Science and Optoelectronics Engineering \& CAS Center of Excellence in Topological Quantum Computation, University of Chinese Academy of Sciences, 100049 Beijing, China}
 \author{Jia-Wei Mei}
 \email{meijw@sustech.edu.cn}
 \affiliation{Shenzhen Institute for Quantum Science and Engineering, and
   Department of Physics, Southern University of Science and Technology,
   Shenzhen 518055, China}
 \affiliation{Shenzhen Key Laboratory of Advanced Quantum Functional Materials
   and Devices, Southern University of Science and Technology, Shenzhen 518055, China}

\date{\today}

\begin{abstract}
Quantum spin liquid (QSL) represents a new class of condensed matter states characterized by the
long-range many-body entanglement of topological orders. The most prominent
feature of the elusive QSL state is the existence of fractionalized spin
excitations. Subject to the strong quantum fluctuations, the spin-1/2
antiferromagnetic system  on a kagome lattice is the promising candidate for
hosting a QSL ground state, but the
structurally ideal realization is rare. Here, we report Raman scattering
measurements on the single crystalline \ce{Cu3Zn(OH)6FBr},  and confirm that 
the ideal kagome structure remains down to low temperatures without any lattice distortion by the angle-resolved
polarized Raman responses and second-harmonic-generation measurements.
Furthermore, at low temperatures the Raman scattering
reveals a continuum of the spin excitations in \ce{Cu3Zn(OH)6FBr}, in
contrast to the  sharp magnon peak in the ordered kagome
antiferromagnet \ce{EuCu3(OH)6Cl3}. Such magnetic Raman continuum,
in particular, the substantial low-energy one-pair spinon excitation serves as strong evidence
for fractionalized spin excitations  in \ce{Cu3Zn(OH)6FBr}.  
\end{abstract}
\maketitle

\section{Introduction}
When subject to strong quantum fluctuation and geometrical frustration, the
quantum spin system may not develop into a magnetically ordered state\cite{Anderson1973,Anderson1987}, but a
quantum spin liquid (QSL) ground state at zero
temperature.\cite{Lee2008,Balents2010,Zhou2017,Broholm2020} QSL has no classic
counterpart as it exhibits various
topological orders characterized by the long-range entanglement
pattern.\cite{Kitaev2006a,Levin2006,Wen2019}
 The lattice
of the spin-1/2 kagome network of corner-sharing triangles is a long-sought platform for
 antiferromagnetically interacting spins to
host a QSL ground state.\cite{Ran2007,Hermele2008,Yan2011,Jiang2012,Zaletel2015,Mei2017,Norman2016} 
Herbersmithite [\ce{ZnCu3(OH)6Cl2}] is the
first promising kagome QSL candidate,\cite{Shores2005,Mendels2007,Lee2008,Wulferding2010,Han2012,Fu2015,Han2016,Norman2016,Khuntia2020} in which no long-range magnetic order
was detected down to low temperature,\cite{Shores2005,Mendels2007} and inelastic neutron scattering on single
crystals revealed a magnetic continuum, as a hallmark of fractionalized spin 
excitations.\cite{Han2012,Han2016} Up to date, most, if not all, experimental information on the
nature of kagome QSL relies on a single compound of Herbertsmithite.
Considering the fact that a lattice distortion away from a perfect kagome structure
has recently been confirmed in Herbersmithite,\cite{Zorko2017,Laurita2019} which stimulates investigations on
the subtle magneto-elastic effect in the kagome
materials,\cite{Norman2020,Li2020a} an alternative realization of the QSL
compound with the ideal kagome lattice is still in urgent need. Zn-Barlowite
[\ce{Cu3Zn(OH)6FBr}] is such a candidate for a kagome
QSL ground state~\cite{Han2014,Liu2015b,Feng2017,Feng2018,Wei2017,Pasco2018,Smaha2018,Smaha2020,Henderson2019,Tustain2020,Wei2020}
with no lattice distortion being reported yet.
Measurements on the powder samples of Zn-Barlowite indicate the absence of
long-range magnetic order or spin freezing down to temperatures of 0.02~K, four
orders of magnitude lower than the Curie-Weiss temperature.\cite{Feng2017,Wei2017} 
Besides the absence of long-range magnetic
order down to low temperature, the fractionalized spin excitations, i.e.
spinons, in the spectroscopy 
is essential evidence for the long-range entanglement
pattern in QSL. However, spectroscopic evidence for the
deconfined spinon excitations in Zn-Barlowite is still lacking, in part due to
unavailable single-crystal samples. Note that for Zn-Barlowite \ce{Cu_{4-x}Zn_{x}(OH)6FBr}
the doping parameter $x\leq0.56$, in the previously reported Zn-Barlowite single
crystal samples, does not belong to the QSL
regime.\cite{Pasco2018,Smaha2018,Smaha2020,Tustain2020,Wei2020}

In this work, we report the synthesis of the single crystals of
\ce{Cu_{4-x}Zn_{x}(OH)6FBr} ($x=0.82$) of millimeter size, which is in the QSL
regime, and the spin
dynamics revealed by the inelastic light scattering on these samples.  We
confirm the ideal kagome-lattice structure by the angle-resolved polarized Raman responses and 
second-harmonic-generation (SHG), and observe a magnetic Raman
continuum in our crystal samples.  Raman scattering has previously been reported for
Herbertsmithite,\cite{Wulferding2010}
and the overall continuum agreed with that in Zn-Barlowite.
Although it was not discussed, the lattice distortion in Herbertsmithite was
evident by the anisotropic angle dependent Raman responses~\cite{Wulferding2010} and 
may account for the difference from our results in details. 
In the theory the Raman spectrum of the kagome QSL
 contains the one-pair component of
spinon-antispinon excitations with a peculiar
power-law behavior at low frequency, serving as the
fingerprint of spinons.\cite{Ko2010} Our measured magnetic Raman continuum agrees
well with the theoretical prediction, revealing the fractionalized spin excitation in
\ce{Cu_{3.18}Zn_{0.82}(OH)6FBr}.

\begin{figure*}[t]
  \centering
  \includegraphics[width=\columnwidth]{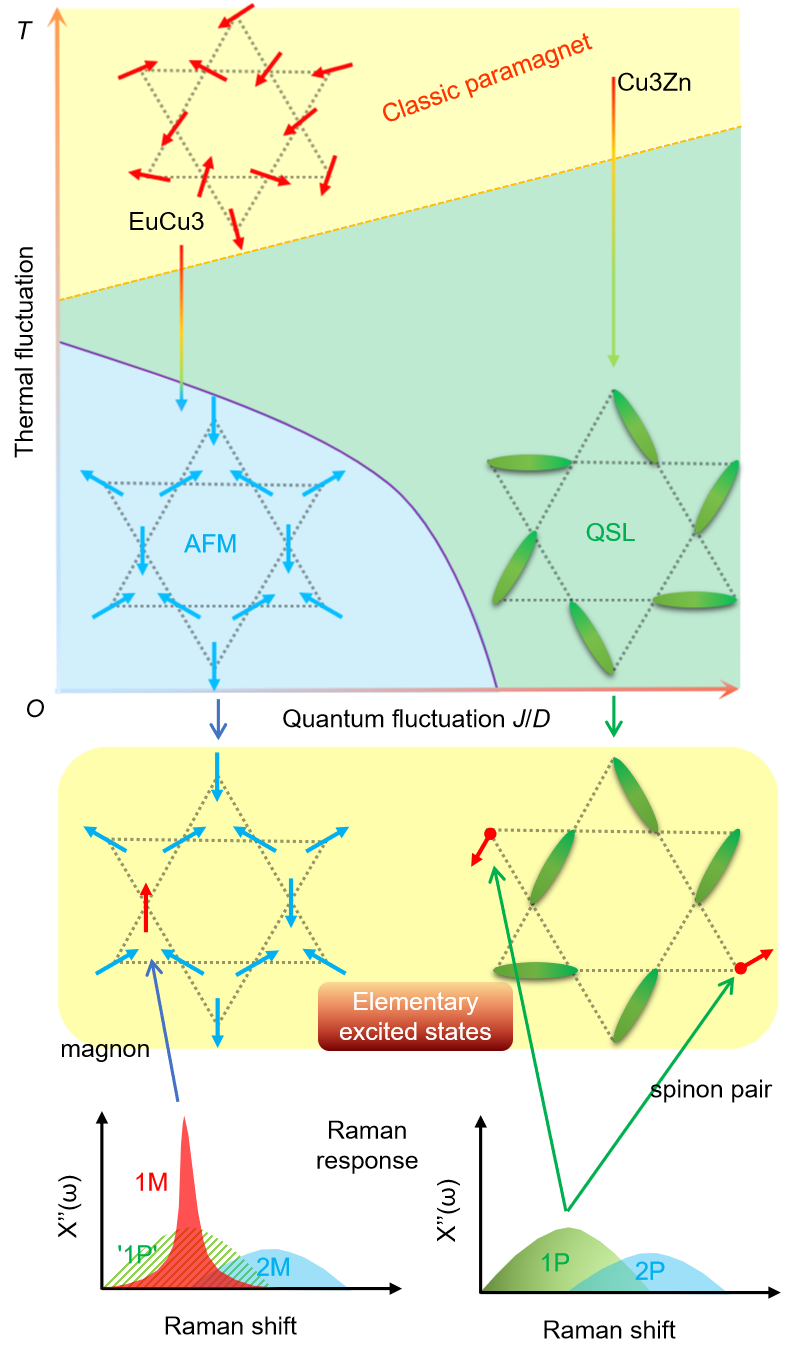}
  \caption{
{\bf Schematical comparative Raman responses for the AFM and QSL states.} With a
large DM interaction $D$, the kagome antiferromagnet develops a 120 $^{\circ}$
non-collinear AFM ground state with the wave vector $\mathbf{q}=0$ below $T_N$.\cite{Cepas2008a,Rousochatzakis2009,Zhu2019} Increasing $J/D$, the
fluctuation of the kagome system increases,  driving the system into the QSL
state. By increasing the temperature, the thermal fluctuation
melts the magnetic order and turns the system into the classic paramagnetic
state at high temperatures. By the first-principle calculations in Supplementary
Section8, Cu3Zn and EuCu3 have the values of $D/J$ as 0.05 and 0.3, and thus
correspond to the
QSL and AFM ground states, respectively. In the middle, the elementary excited states of AFM
and QSL states are the magnon and spinon, respectively, resulting in different magnetic Raman
spectra shown at the bottom. Here 1P and 2P denote the one-pair and two-pair
spinon excitations, respectively. 1M and 2M in magnetically ordered state denote
the one- and two-magnon excitations, respectively. The 1M Raman peak in AFM
measures the magnon [Methods Section] while the 1P Raman continuum in QSL probes
the spinon excitations.\cite{Ko2010} The shadow background of the 1M peak,
marked as `1P', denotes the continuum above $T_N$ in EuCu3, mimicking the 1P
continuum in the QSL state [Supplementary Section6]. So the magnon excitation
below $T_N$ emerges from the one-pair continuum and can be regarded as the bound
state of the spinon-antispinon excitations. The transition between QSL and AFM
can be thought to be driven by the spinon confinement.} 
  \label{fig:figure1}
\end{figure*}

To demonstrate the one-pair spinon dynamics in the kagome QSL even
more evidently, we perform a control
experiment on a kagome antiferromagnet \ce{EuCu3(OH)6Cl3}, which suffers a
spinon confinement as a transition taking places from a paramagnetic phase to a
$\mathbf{q}=0$ type
 120$^\circ$ non-collinear  antiferromagnetic  order (AFM) ground state below
the N\'eel temperature $T_N=17$~K.\cite{Puphal2018,Zorko2019,Arh2020} We observe a
magnon peak in the AFM state,   which can be regarded as the spinon confinement
in the magnetically ordered state as schematically summarized in
Fig.~\ref{fig:figure1}. The magnon excitation emerges from
the one-pair continuum, firstly reported in our
work, and can be regarded as the bound state of the
spinon-antispinon excitations.

\section{Results}
We synthesized single crystals of Barlowite \ce{Cu4(OH)6FBr}, Zn-Barlowite
\ce{Cu_{3.18}Zn_{0.82}(OH)6FBr}, and \ce{EuCu3(OH)6Cl3} (we use the
short-hand notation Cu4, Cu3Zn, and EuCu3, respectively) with high quality
(Methods and Supplementary Section1). The crystal structure of Cu3Zn was
determined by the single-crystal X-ray diffraction, and chemical analysis
(ICP-AES) indicated the crystal composition is Cu$_{3.18}$Zn$_{0.82}$(OH)$_6$FBr
(Methods), in the QSL region of the magnetic phase
diagram.\cite{Wei2020,Tustain2020} The interlayer Cu$^{2+}$ concentration (18\%) is
comparable to that (15\%) in Herbertsmithite.\cite{Freedman2010} The crystal structure
of Cu4 coincides with Barlowite \textbf{1} in Refs.~\cite{Smaha2018,Smaha2020}.
We fit the high-temperature magnetic susceptibility (Supplementary Section1) by
the Curie-Weiss behavior to estimate the superexchange strength for the
Cu$^{2+}$ spin-1/2 moments on the kagome lattice in Cu3Zn, $J\simeq19$~meV
($\Theta_{\text{CW}}=-220$~K). The antiferromagnetic super-exchange interaction
for EuCu3 is about $J\simeq10$~meV.\cite{Puphal2018,Zorko2019,Arh2020} Note the
electronic ground state of Eu$^{3+}$ in EuCu3 is the non-magnetic $^7F_0$
configuration.

To study changes in the crystal structures of Cu3Zn and Cu4, we track the
temperature evolution of Raman spectra in the two compounds. Cu3Zn and Cu4 at high temperature crystallize the same space group
$P6_3/mmc$.\cite{Feng2017,Han2014} We didn't observe the structural phase
transition in Cu3Zn from the Raman scattering down to low temperature
(Supplementary Section2 and 3). Cu4 transforms to orthorhombic \textit{Pnma}
below 265~K, characterized by changes in the relative occupancies of the
interlayer Cu$^{2+}$ site.\cite{Feng2018,Pasco2018,Smaha2018,Smaha2020} The
splitting of phonon peaks in Cu4 due to the superlattice folding in the
orthorhombic Pnma phase is resolved for several modes [Supplementary Fig.S4]. Cu3Zn displays sharp
$E_{2g}$ modes at 125~cm$^{-1}$ for in-plane relative movements of Zn$^{2+}$.
The corresponding modes for the interlayer Cu$^{2+}$ in Cu4 are broad at 290~K
due to the randomly distributed interlayer Cu$^{2+}$ and split into two peaks
below the structural transition temperature. Cu3Zn has no Raman-active mode related to the
kagome Cu$^{2+}$ vibrations, indicating the kagome layer remains substantially
intact as the inversion center of Cu$^{2+}$ sites is evident. The kagome
layers in Cu4 are distorted, signaled by a new phonon mode for the kagome
Cu$^{2+}$ vibration at 62~cm$^{-1}$. Besides sharp phonon modes, we observe a
Raman continuum background in Cu3Zn, particularly at low frequency, signifying
substantial magnetic excitations.

Previous X-ray and neutron refinements of the crystal structure suggest ideal kagome
planes in Cu3Zn. SHG confirmed the parity symmetry of the crystallographic structure in
 Barlowite {\bf 2} \ce{Cu4(OH)6FBr} and Zn-Barlowite
\ce{Cu_{3.66}Zn_{0.33}(OH)6FBr}. In Supplementary
Section7, we also reveal the inversion symmetry by SHG in our single crystals
of Cu3Zn. 
To further exclude subtle local symmetry lowering or lattice
distortions, we perform the angle-resolved polarization-dependent Raman
measurements of Cu3Zn for magnetic and lattice vibration
modes.\cite{Cepas2008,Liu2017}
The threefold rotation symmetry of the kagome lattice leads to 
isotropic angle
dependence in the $XX$ configuration both for $A_{1g}$ and $E_{2g}$ components, $XY$
and $X$-only configurations
for the $E_{2g}$ component; it also gives rise to the angle $\theta$ dependence of $\cos^2\theta$
in the $X$-only configuration for the $A_{1g}$ component. 
We find that the angle dependence of Raman
responses, in particular for the magnetic continuum at low frequency, the Br$^-$ $E_{2g}$
phonon, and O$^{2-}$ $A_{1g}$ phonon modes, fit the theoretical curves very well
[Supplementary Section4], confirming the threefold rotational symmetry of the
kagome lattice in the dynamical Raman responses of the lattice vibrations and
magnetic excitations. Combining the  X-ray and neutron
refinement,\cite{Feng2017,Feng2018,Wei2017,Wei2020} 
we conclude that Cu3Zn manifests a structurally ideal realization of
layered spin-1/2 Cu$^{2+}$ kagome-lattice planes.

\begin{figure*}[t]
\centering
\includegraphics[width=1.5\columnwidth]{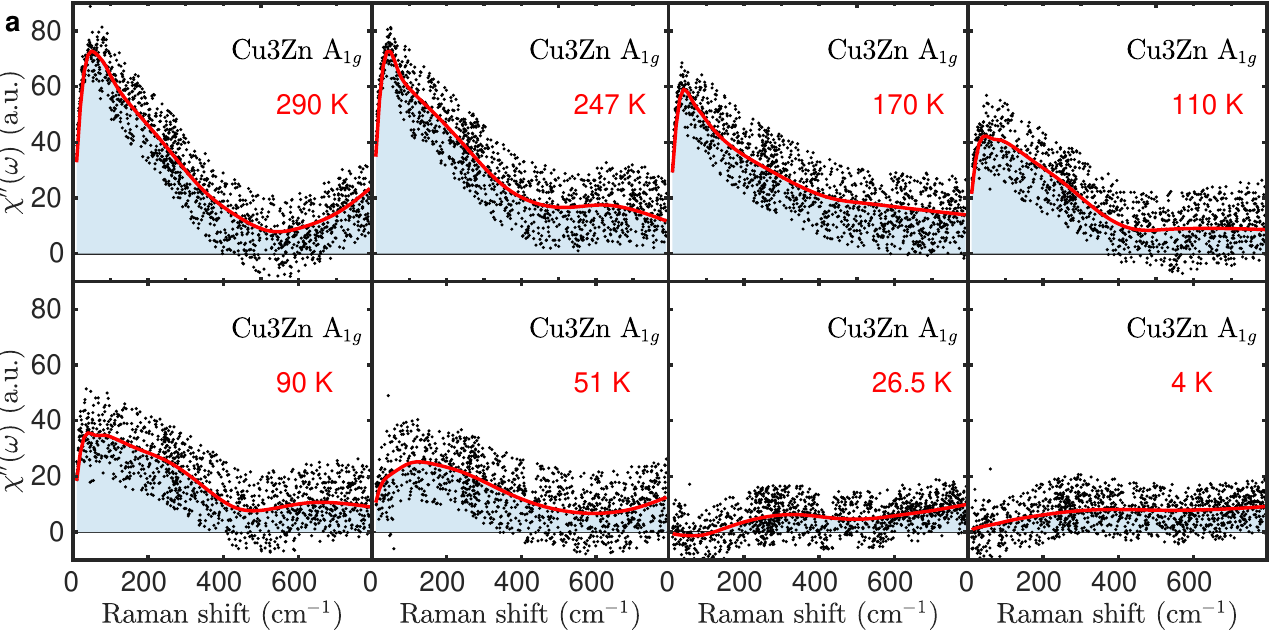} \includegraphics[width=0.5\columnwidth]{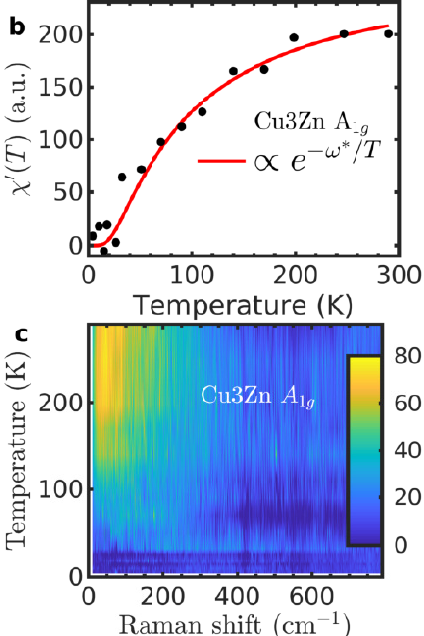}
\includegraphics[width=1.5\columnwidth]{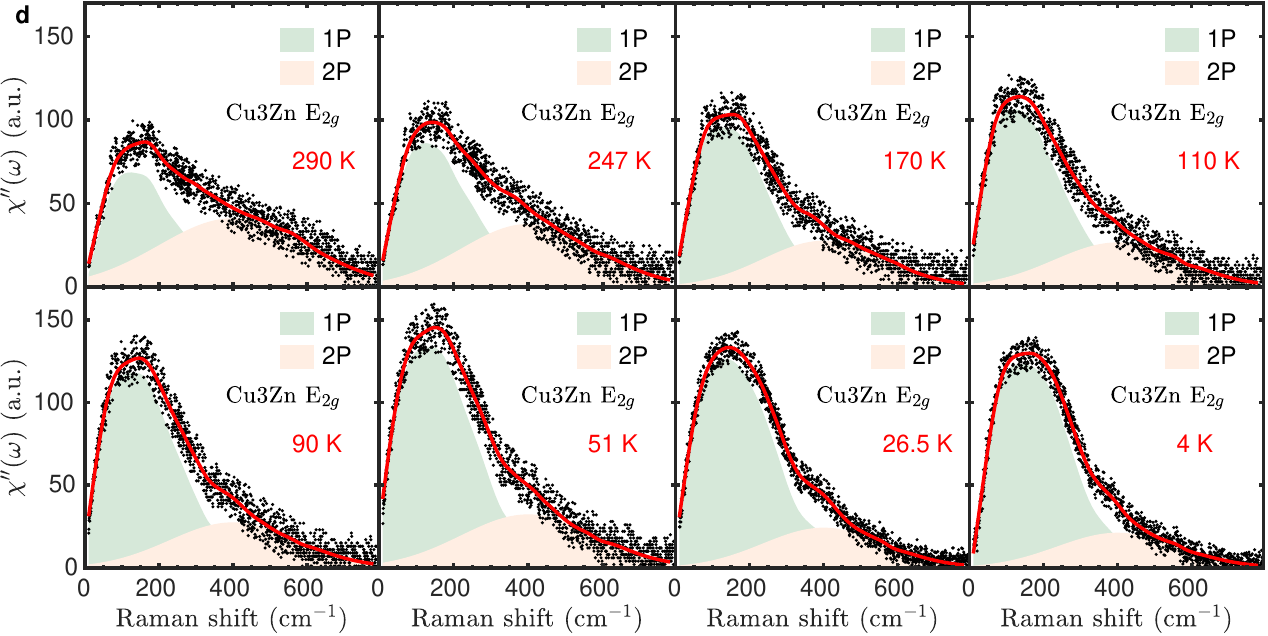} \includegraphics[width=0.5\columnwidth]{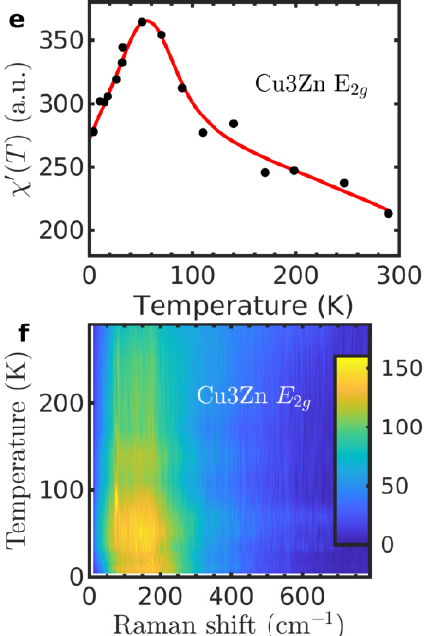}
\caption{
{\bf Temperature dependence of the Raman susceptibilities  in Cu3Zn. } ({\bf a}) The $A_{1g}$ Raman susceptibility $\chi''_{A_{1g}}=\chi''_{XX}-\chi''_{XY}$. The solid lines are a guide to the eye. ({\bf b}) Temperature dependence of the static Raman susceptibility in $A_{1g}$ channel $\chi_{A_{1g}}'(T)=\frac{2}{\pi}\int_{10}^{400\text{~cm}^{-1}}\frac{\chi''_{A_{1g}}(\omega)}{\omega}d\omega$. The solid line is a thermally activated function. ({\bf c}) Color map of the temperature dependence of the magnetic Raman continuum $\chi''_{A_{1g}}(\omega,T)$. ({\bf d}) The $E_{2g}$ Raman response function $\chi''_{E_{2g}}=\chi'_{XY}$. The solid lines are a guide to the eye. We schematically decompose the $E_{2g}$ magnetic Raman continuum into two components of spin excitations for one and two spin-antispinon pair excitations, respectively. Here 1P and 2P represent one- and two-pair, respectively. ({\bf e}) Temperature dependence of the static Raman susceptibility in $E_{2g}$ channel $\chi_{E_{2g}}'=\frac{2}{\pi}\int_{10}^{780\text{~cm}^{-1}}\frac{\chi''_{E_{2g}}(\omega)}{\omega}d\omega$. The solid line is a guide to the eye.  ({\bf f}) Color map of the temperature dependence of the magnetic Raman continuum $\chi''_{E_{2g}}(\omega,T)$.
}
  \label{fig:figure2}
\end{figure*}

Having established the absence of a sharp anomaly in the thermodynamic
properties [Supplementary Section1] and the lack of an emergent magnetic order
with the weak symmetry breaking in the angle-dependent polarized Raman response, which is
the first step to a QSL, we now present our main results of the magnetic Raman
continuum in Cu3Zn with subtracting phonon contribution, as shown  in Fig.~\ref{fig:figure2}. The susceptibility is related to
the Raman intensity $I(\omega)=(1+n(\omega))\chi''(\omega)$ with the bosonic
temperature factor $n(\omega)$. Fig.~\ref{fig:figure2}a, b and c are the
$A_{1g}$ magnetic Raman response in Cu3Zn, which  measures the thermal
fluctuation of the interacting spins on the kagome
lattice.\cite{Halley1978,Reiter1976,Lemmens2003} We can see that the $A_{1g}$
channel is activated only at high temperatures, disappears at low temperatures,
behaving as the thermally activated excitations. At high temperatures, the Raman
spectra exhibit the quasielastic scattering that is common in the inelastic
light scattering for the spin systems.\cite{Lemmens2003} The maximum in the
Raman response function decreases from 60~cm$^{-1}$  at room temperature to
30~cm$^{-1}$ at 110~K, and the magnetic intensity becomes hardly resolved at low
temperatures below 50~K. The integrated Raman susceptibility $\chi'(T)$ in
Fig.~\ref{fig:figure2}b fits the thermally activated function, $\propto
e^{-\omega^*/T}$ with $\omega^*=53$~cm$^{-1}$, different from the power-law
temperature dependence of the quasielastic scattering in
Herbertsmithite.\cite{Wulferding2010} The temperature dependence of the $A_{1g}$
magnetic Raman susceptibilities $\chi_{A_{1g}}''(\omega,T)$ in Cu3Zn distributes
the main spectral weight among the frequency region less than 400~cm$^{-1}$ and
the temperature range above 50~K in Fig.~\ref{fig:figure2}c. 

Different from the $A_{1g}$ channel, the pronounced $E_{2g}$ magnetic Raman
continuum in Cu3Zn persists at low temperatures
(Fig.~\ref{fig:figure2}d), indicating the quantum fluctuation of the kagome
spin-1/2 system. Along with the theoretical work,\cite{Ko2010} we schematically
decompose the $E_{2g}$ Raman continuum into two components, which have
the maximum around 150~cm$^{-1}$ and the higher one around 400~cm$^{-1}$, respectively. We
denote them as spin excitations for one spin-antispinon pair (one-pair) and two spinon-antispinon pair (two-pair) excitations,
respectively.\cite{Ko2010} Like the two-magnon scattering in the
antiferromagnet,\cite{Lemmens2003,Devereaux2007} the two-pair component doesn't
show a significantly non-monotonic temperature dependence as reducing
temperatures. The substantial low energy one-pair component has a more
pronounced non-monotonic temperature dependence. It increases with the
temperature decreasing from 290~K to 50~K and decreases with further temperature
reduce as shown in Figs.~\ref{fig:figure2}d, e, and f. The frequency and temperature
dependence of the $E_{2g}$ magnetic Raman susceptibilities
$\chi_{E_{2g}}''(\omega,T)$ distributes  the main spectral weight among the
frequency region less than 400~cm$^{-1}$, and reaches the maximum at around
150~cm$^{-1}$ and 50~K,
as shown in Fig.~\ref{fig:figure2}f. We also observed the Fano effect for the
$E_{2g}$ F$^-$ phonon peak at 173~cm$^{-1}$ in Cu3Zn  [Supplementary Section3],
whose asymmetric lineshape provides an additional probe of the magnetic
continuum.

The one-pair component in the $E_{2g}$ Raman continuum is crucial as it has an
origin in the spinon excitation in the kagome QSL from the perspective of theory.\cite{Ko2010} With the
incoming and outgoing light polarizations $\hat{\mathbf{e}}_{\text{in}}$ and
$\hat{\mathbf{e}}_{\text{out}}$,  magnetic Raman scattering measures the
spin-pair (two-spin-flip) dynamics in terms of the Raman tensor\cite{Fleury1968,Shastry1990,Cepas2008,Ko2010}
\begin{eqnarray}
  \label{eq:cross_sec}
  \tau_R=\sum_{\langle ij
  \rangle}(\hat{\mathbf{e}}_{\text{in}}\cdot\mathbf{r}_{ij})(\hat{\mathbf{e}}_{\text{out}}\cdot
  \mathbf{r}_{ij})\mathbf{S}_i\cdot\mathbf{S}_j,
\end{eqnarray}
where the summation runs over $\langle ij \rangle$ for the nearest neighbor
bonds $\mathbf{r}_{ij}$ for the $\mathbf{S}_i$ and $\mathbf{S}_j$ on the kagome
lattice. At zero temperature, the magnetic Raman susceptibility is given as
$\chi''_R(\omega)=\sum_f|\langle f|\tau_R|0
\rangle|^2\delta(\omega-\omega_f))=\sum_f\mathcal{M}^R_{f}(\omega_f)\delta(\omega-\omega_f)\simeq
M(\omega)\mathcal{D}(\omega)$, in which we set the ground state energy to be
zero. $\mathcal{M}^R_{f}(\omega_f)=|\langle f|\tau_R|0\rangle|^2$ denotes the
matrix element for the transition between the ground state $|0\rangle$ and the
excited state $|f\rangle$.
$M(\omega)=\sum_{f:\omega_f=\omega}\mathcal{M}_f^R=\overline{\mathcal{M}}_f^R(\omega_f)$
is the average matrix element, and $\mathcal{D}(\omega)=\sum_{f:\tau_R\text{
    allowed final states~}}\delta(\omega-\omega_f)$ denotes the density of state
(DOS) for the Raman tensor associated excitations. Introducing the spinon operator
$f_{i\sigma}$ in QSL, the spin-pair operator in the Raman tensor is rewritten
in terms of two pairs of spinon-antispinon excitations $\tau_R\propto
f_{i\sigma}^\dag f_{j\sigma} f_{j\sigma'}^\dag f_{i\sigma'}$. Besides the
two-pair excitation, the magnetic Raman continuum contains the one-pair spinon
excitation as $\tau_R\propto \chi (f_{i\sigma}^\dag f_{j\sigma}+f_{j\sigma}^\dag
f_{i\sigma})$, where $\chi=\langle f_{i\sigma}^\dag f_{j\sigma}  \rangle$ is the
spinon mean-field hopping amplitude.\cite{Ko2010} As shown in
Fig.~\ref{fig:figure2}d, the one-pair component in the measured Raman
susceptibility in the $E_{2g}$  has the maximum at 150~cm$^{-1}$ ($J$), and
extends up to 400~cm$^{-1}$ ($2.6J$) at low temperatures. The two-pair component
has the maximum at 400~cm$^{-1}$ (2.6~$J$) and the cut-off around 750~cm$^{-1}$
(4.9~$J$). In totality, the mentioned features (maxima and cut-offs) of one- and two-pair excitations in the $E_{2g}$ measured
Raman response in Cu3Zn (Fig.~\ref{fig:figure2} d) overall agree well with the
theoretical calculation for the kagome QSL state.\cite{Ko2010}

\begin{figure*}[t]
\centering
\includegraphics[width=1.8\columnwidth]{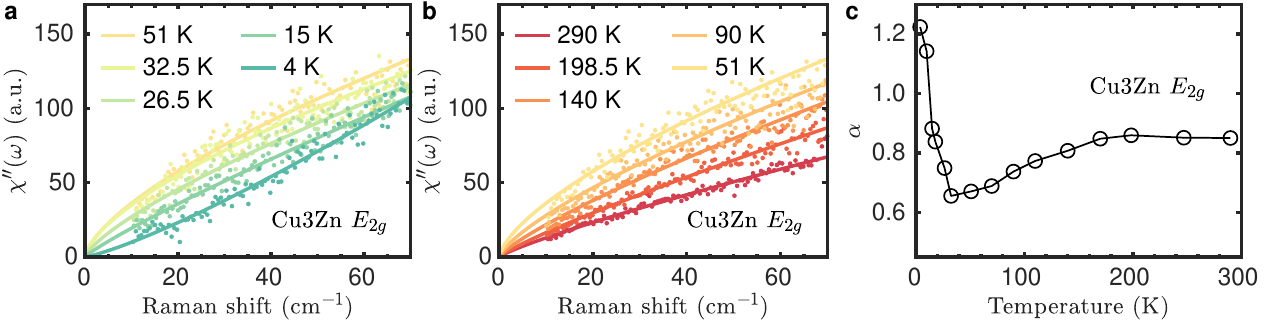}
\caption{\textbf{Power-law behavior for the $E_{2g}$ Raman continua at low frequency in
    Cu3Zn.}  ({\bf a}) and ({\bf b}) Power-law fitting of $\chi''_{E_{2g}}(\omega)\propto\omega^\alpha$
  at low and high temperatures, respectively, in Cu3Zn.  ({\bf c})
    Temperature dependent  exponent $\alpha$ for the power-law fittings in cu3zn.
}
  \label{fig:figure3}
\end{figure*}

In more detail, the one-pair component  dominates the $E_{2g}$ magnetic Raman continuum at low
frequency. It displays the power-law behavior up to 70~cm$^{-1}$, with a
significantly nonmonotonic temperature dependence, as shown in
Figs.~\ref{fig:figure3}. As
lowering the temperature, the $E_{2g}$ continuum at low frequency increases
above 50~K and decreases below 50~K. The low-energy continuum evolves from a
sublinear behavior
$T^{\alpha}$ with $\alpha<1$ to a superlinear one $T^{\alpha}$ with
$\alpha>1$ as reducing the temperature. 
A central question for the kagome spin liquid is whether a
spin gap exist. The results of the spin gap in Herbertsmithite
are controversial due to the difficulty of singling out the kagome
susceptibility.\cite{Fu2015,Khuntia2020} Previous results on the powder
samples of Cu3Zn suggest a small spin gap,\cite{Feng2017,Wei2017} and
measurements on the single-crystal samples would  be of great interest. If such a gap
exists, the power-law behavior of the $E_{2g}$ magnetic Raman continua  sets  an
upper bound for the spin gap of 2~meV.

The temperature-dependent magnetic continua of Cu3Zn in
Figs.~\ref{fig:figure2}d, e, and f, and Fig.~\ref{fig:figure3} imply the
maximal spin fluctuations at the characteristic temperature 50~K. The
maximum of the kagome spin fluctuations in Cu3Zn signifies the spin singlet
forming,\cite{Anderson1987,Lee2006} but is masked by the inter-layer Cu$^{2+}$ moments in the thermodynamic
properties [Supplementary Section1]. It can be revealed by the Knight shift
related to the kagome spins in the nuclear magnetic resonace
measurements.\cite{Feng2017}  In contrast to
significant energy dependence in magnetic Raman continuum in Cu3Zn in
Figs.~\ref{fig:figure2} and~\ref{fig:figure3}, the scattered neutron signal in
Herbertsmithite is overall insensitive to energy transfer, rather flat above
1.5~meV, but increases significantly at low-energy scattering due to the
interlayer Cu$^{2+}$ ions.\cite{Han2012,Han2016} The interlayer Cu$^{2+}$ ions
distribute spatially away from each other, and the spin-pair magnitude among
themselves and between them and kagome spins  is weak, giving rise to a negligible
matrix element in Raman tensors. So different from the neutron scattering, the
Raman scattering is not sensitive to the inter-layer Cu$^{2+}$ at
low energies, advantageous to the detection of kagome spins.
Furthermore, inelastic neutron scattering in
Herbertsmithite measures the magnetic continuum up to 2-3~$J$,\cite{Han2012} the
same energy range  as the one-pair Raman component in Cu3Zn. These results suggest that the magnetic Raman
continuum originates from the kagome-plane spins, and the one-pair component has an
origin of spinon excitations.

The theoretical calculation for kagome Dirac
spin liquid (DSL) predicts the power-law behavior for the Raman susceptibility
in the $E_{2g}$ channel at low frequency.\cite{Ko2010} The one-pair spinon
excitation in DSL gives the linear density of state
$\mathcal{D}_{\text{1P}}\propto \omega$.  The matrix element turns out to be
exactly zero for all one-pair excitations in the mean field Dirac Hamiltonian.
As a result a Raman spectrum that scales as $\omega^3$ was
predicted.\cite{Ko2010} However, the vanishing of the matrix element is somewhat
accidental and depends on the assumption of a DSL in a Heisenberg model in an
ideal kagome lattice. Any deviation from the ideal DSL state, \textit{e.g.} a
small gap in the ground state,\cite{Feng2017,Wei2017} DM interactions, or other
effects of perturbations,\cite{Bernu2020,Norman2020} changes the wave functions
and may result in a constant matrix element $M(\omega)$. In that case, the Raman
spectrum will be simply proportional to the DOS of the one-pair component
$\mathcal{D}_{\text{1P}}$ which is linear in $\omega$. From our fitting for Cu3Zn
in Fig.~\ref{fig:figure3}, we find that $\alpha=1.3$ when approaching zero
temperature. The existence of a small gap in the spinon spectrum may explain
this discrepancy. We also note that according to the theory~\cite{Ko2010} the
$A_{1g}$ and $A_{2g}$ contributions to the one-pair continuum are the
forth-order effect, much smaller than
the $E_{2g}$ contributions. This explains the invisible one-pair continuum in the
$A_{1g}$  and $A_{2g}$ channels.   

\begin{figure*}[t]
\centering
\includegraphics[width=1.5\columnwidth]{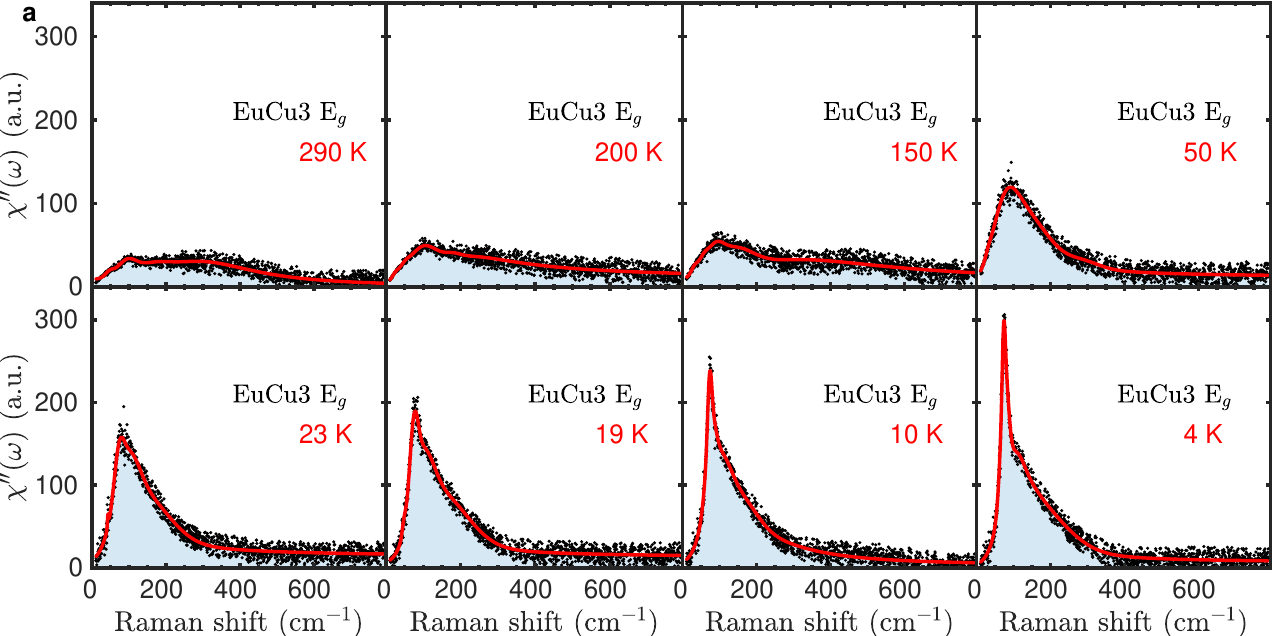} \includegraphics[width=0.5\columnwidth]{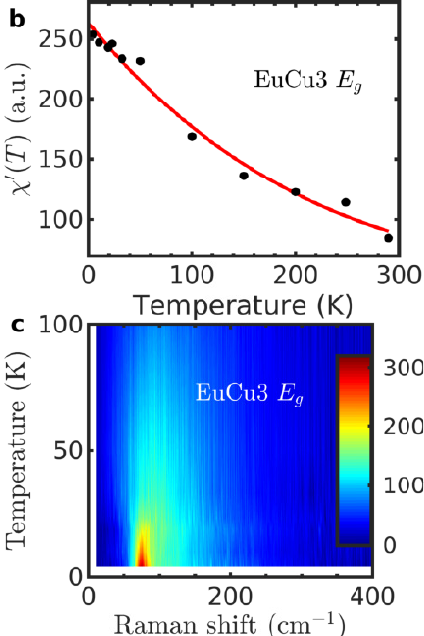}
 \caption{{\bf temperature dependence of the $e_{g}$ raman susceptibilities
     in
     eucu3. } 
   ({\bf a}) the $e_{g}$ raman susceptibility  $\chi''_{e_{g}}=\chi''_{xy}$. the solid lines are a guide to
     the eye. a sharp magnon peak appears in the $e_{g}$ magnetic raman continuum below the magnetic transition temperature
$t_n=17$~k.  ({\bf b}) temperature
dependence of the static raman susceptibility in $e_{g}$ channel
$\chi_{e_{g}}'=\frac{2}{\pi}\int_{10}^{780\text{~cm}^{-1}}\frac{\chi''_{e_{g}}(\omega)}{\omega}d\omega$.
the solid line is a guide to the eye.  ({\bf c}) Color map of the temperature
dependence of the magnetic Raman continuum $\chi''_{E_{g}}(\omega,T)$. A
sharp magnon peak is observed below $T_N$.
}
  \label{fig:figure4}
\end{figure*}

Figure~\ref{fig:figure4} presents a control Raman study on the magnetic ordered
kagome antiferromagnet EuCu3, which has the antiferromagnetic superexchange
$J\simeq10$~meV, half of the value in Cu3Zn. EuCu3 belongs to the atacamite
family with the perfect kagome lattice and has the $\mathbf{q}=0$ type 120$^\circ$
ordered spin configuration below $T_N$ due to a large Dzyaloshinski-Moriya (DM)
interaction [Supplementary
Section8].\cite{Sun2016,Okuma2017,Puphal2018,Iida2020,Arh2020} Above the
ordering temperature $T_N=17$~K, the magnetic Raman continuum in the $E_{g}$
channel displays the extended continuum, similar to that in Zn-Barlowite at 4~K as shown
in Fig.S10 in Supplementary Section6. This indicates the strong magnetic
fluctuations in EuCu3. The less pronounced low-energy continuum
excitations in \ce{EuCu3(OH)6Cl3} indicate the suppression of the quantum
fluctuation due to a large DM interaction. The low energy excitation in the ordered state is the
spin-wave excitation, i.e.\  magnon, and the $E_{g}$ Raman scattering measures
one- and two-magnon excitations for the non-collinear 120$^\circ$ spin
configuration as detailed in the Methods section, leading to a sharp magnon peak
at 72~cm$^{-1}$ superimposing on the two-magnon continuum. In this sense, the AFM transition may be thought of as a confinement transition. The comparative
studies between Cu3Zn and EuCu3 are sketched in Fig.~\ref{fig:figure1},
demonstrating the spinon deconfinement and confinement, respectively, in the
different ground states.

\section{Conclusions}

Our Raman scattering studies  compare the spin dynamics in the kagome QSL
compound Cu3Zn and magnetically ordered antiferromagnet EuCu3. In contrast to a
sharp magnon peak in EuCu3, the overall magnetic Raman scattering in Cu3Zn agrees well
with the theoretical prediction for a spin liquid state. The spinon
continuum is evident, providing the strongest evidence yet for the kagome QSL
ground state in Cu3Zn. On the material side, Zn-Barlowite provides
an ideal structural realization of the kagome lattice, and the available
single crystal samples stimulate future systematical studies of the kagome QSL. Along with Herbertsmithite, the single-crystalline
Zn-Barlowite stands able to provide considerable insight into singling out the
intrinsic properties of the intrinsic nature of the kagome QSL, without
deceiving by the material chemistry details.  

\bibliography{../Refs}

 \section*{Methods}
 {\bf Sample preparation and characterization.}
High qualified single crystals of Zn-Barlowite was
grown by a hydrothermal method similar to crystal growth of
herbertsmithite.\cite{Chu2011,Velazquez2020} \ce{CuO} (0.6 g), \ce{ZnBr2} (3 g), and
\ce{NH4F} (0.5 g) and 18 ml deionized water were sealed in a quartz tube and
heated between 200~$^{\circ}$C and 140~$^{\circ}$C by a two-zone furnace. After three
months, we obtained millimeter-sized single crystal samples. The value of $x$ in
\ce{Cu_{4-x}Zn_x(OH)6FBr} has been determined as 0.82 by Inductively Coupled
Plasma-Atomic Emission Spectroscopy (ICP-AES). The single-crystal X-ray
diffraction has been carried out at room temperature by using Cu source
radiation ( $\lambda=1.54178$~\AA) and solved by the Olex2.PC suite
programs.\cite{Dolomanov2009} The structure and cell parameters of
\ce{Cu_{4-x}Zn_x(OH)6FBr} are in coincidence with the previous report on
polycrystalline samples.\cite{Feng2017,Wei2017} For
Barlowite(\ce{Cu4(OH)6FBr}), the mixture of \ce{CuO} (0.6 g), \ce{MgBr2} (1.2 g), and
\ce{NH4F} (0.5 g) was transferred into Teflon-lined autoclave with 10 ml water.
The autoclave was heated up to 260~$^{\circ}$C and cooled to 140~$^{\circ}$C
after two weeks. A similar growth condition to Barlowite was applied for the
growth of \ce{EuCu3(OH)6Cl3} with staring materials of \ce{EuCl3.6H2O}
(2 g) and \ce{CuO} (0.6 g).  

~

{\bf Measurement methods.}  Our thermodynamical 
measurements were carried out on the Physical Properties Measurement System
(PPMS, Quantum Design) and the Magnetic Property Measurement System (MPMS3,
Quantum Design).

The temperature-dependent Raman spectra are measured in a backscattering geometry using a home-modified Jobin-Yvon HR800 Raman system equipped with an electron-multiplying charged-coupled detector (CCD) and a 50$\times$ objective with long working distance and numerical aperture of 0.45. The laser excitation wavelength is 514~nm from an Ar$^+$ laser. The laser-plasma lines are removed using a BragGrate bandpass filter (OptiGrate Corp.), while the Rayleigh line is suppressed using three BragGrate notch filters (BNFs) with an optical density 4 and a spectral bandwidth $\sim$5-10~cm$^{-1}$. The 1800 lines/mm grating enables each CCD pixel to cover 0.6 cm$^{-1}$. The samples are cooled down to 30 K using a Montana cryostat system under a vacuum of 0.4 mTorr and down to 4 K using an attoDRY 1000 cryogenic system. All the measurements are performed with a laser power below 1 mW to avoid sample heating. The temperature is calibrated by the Stocks-anti-Stocks relation for the magnetic Raman continuum and phonon peaks. The intensities in two cryostat systems are matched by the Raman susceptibility. 
The polarized Raman measurements with light polarized in the $ab$ kagome plane of samples were performed in
parallel ($XX$), perpendicular ($XY$), and $X$-only  polarization configurations
[Supplementary Section4].

SHG measurements were performed using a homemade confocal microscope in a
back-scattering geometry. A fundamental wave centered at 800 nm was used as
excitation source, which was generated from a Ti-sapphire oscillator (Chameleon
Ultra II) with an 80 MHz repetition frequency and a 150 fs pulse width. After
passing through a 50x objective, the pump beam was focused on the sample with a
diameter of 2 $\mu$m. The scattering SHG signals at 400 nm were collected by the
same objective and led to the entrance slit of a spectrometer equipped with a
thermoelectrically cooled CCD. Two shortpass filters were employed to cut the
fundamental wave.

~

{\bf Magnon peak in Raman response for $\mathbf{q}=0$ AFM state.} 
We consider a kagome lattice antiferromagnet with the DM interaction
\begin{eqnarray}
  \label{eq:kdm}
  H=J\sum_{\langle {ij} \rangle}(\mathbf{S}_i\cdot\mathbf{S}_j)+D\hat{z}\cdot\sum_{\langle ij \rangle}\mathbf{S}_i\times\mathbf{S}_j,
\end{eqnarray}
where summation runs over nearest neighbor bonds $\langle {ij} \rangle$ of the kagome lattice,
and the DM interaction is assumed to be of the out-of-plane type. With  a large DM interaction $D$, the kagome antiferromagnet devoleps a
$\mathbf{q}=0$ type 120$^\circ$ AFM order at low temperature in
EuCu3.\cite{Cepas2008a,Rousochatzakis2009,Zhu2019,Puphal2018,Zorko2019,Arh2020}
In terms of the local basis for the AFM order, we rewrite the Hamiltonian as
\begin{eqnarray}
  \label{eq:HlocalR}
H&=&J\sum_{\langle{ij}  \rangle}\mathbf{S}_i\odot\mathbf{S}_j+D\sum_{\langle{ij}  \rangle}\mathbf{S}_i\otimes\mathbf{S}_j,
\end{eqnarray}
with 
\begin{eqnarray}
  \label{eq:spin-pair}
  \mathbf{S}_i\odot\mathbf{S}_j&=&S_i^xS_j^x+\cos(\theta_{ij})(S_i^yS_j^y+S_i^zS_j^z)\nonumber\\
                               &&+\sin(\theta_{ij})(S_i^zS_j^y-S_i^yS_j^z),\nonumber\\
  \mathbf{S}_i\otimes\mathbf{S}_j&=&\sin(\theta_{ij})(S_i^yS_j^y+S_i^zS_j^z)\nonumber\\
  &&+\cos(\theta_{ij})(S_i^yS_j^z-S_i^zS_j^y),
\end{eqnarray}
where $\theta_{ij}$ is an angle between two neighboring spins. The effective linear
spin wave Hamiltonian is given as
\begin{eqnarray}
  \label{eq:SWT}
  \mathcal{H}_{\text{eff}}&=&J\sum_{\langle{ij}
                              \rangle}[S_i^xS_j^x+(\cos\theta_{ij}+\sin\theta_{ij}D/J)\times\nonumber\\
  &&(S_i^yS_j^y+S_i^zS_j^z)],
\end{eqnarray}
for  which the Holstein-Primakoff representation for spin operators
in the local basis was applied and the energy dispersion was obtained in Ref.~\cite{Chernyshev2014}. 

The Raman tensor in the $XY$ configuration is given as
\begin{eqnarray}
  \label{eq:Egtr}
  &&\tau_R^{xy}=\sum_{\langle {ij}
                 \rangle}\mathbf{r}_{ij}^x\mathbf{r}_{ij}^y\mathbf{S}_i\cdot\mathbf{S}_j,\nonumber\\
  &=&\frac{\sqrt{3}}{4}\sum_R\mathbf{S}_{R3}\cdot(\mathbf{S}_{R1}+\mathbf{S}_{R+\mathbf{a}_21}-\mathbf{S}_2-\mathbf{S}_{R-\mathbf{a}_1+\mathbf{a}_22}).
\end{eqnarray}
In the local spin basis, we have the Raman
tensor is given as
\begin{eqnarray}
  \label{eq:tr}
  \tau_R^{xy}=\frac{\sqrt{3}}{4}\sum_R&&\mathbf{S}_{R3}\odot(\mathbf{S}_{R1}+\mathbf{S}_{R+\mathbf{a}_21}-\mathbf{S}_2\nonumber\\
  &&-\mathbf{S}_{R-\mathbf{a}_1+\mathbf{a}_22}).
\end{eqnarray}
In the spin-pair operator $\mathbf{S}_i\odot\mathbf{S}_j$ in Eq.~(\ref{eq:spin-pair}), there are two-magnon contribution
in terms of $S_i^xS_j^x+\cos(\theta_{ij})(S_i^yS_j^y+S_i^zS_j^z)$, and one- and
three-magnon contributions in terms of
$\sin(\theta_{ij})(S_i^zS_j^y-S_i^yS_j^z)$. For the $\mathbf{q}=0$ spin
configuration, we find that $\tau_R^{xy}$ in Eq.~\ref{eq:tr} has the
non-vanished one magnon contributions. For the $\sqrt{3}\times\sqrt{3}$ AFM state,
$\tau_R^{xy}$ has no one-magnon contribution. The observed one-magnon peak in
the $E_{g}$ channel in EuCu$_3$ provides  evidence for the $\mathbf{q}=0$
spin ordering at low temperatures. In the linear spin-wave theory, we take $S^z$
in the local basis as a constant, $S_i^z=\langle S^z \rangle=1/2$, and the $XY$ Raman tensor is given as
\begin{eqnarray}
  \label{eq:trxy}
  \tau_R^{xy}=\frac{3}{8}\sum_R(S_{R1}^y+S_{R2}^y-2S_{R3}^y),
\end{eqnarray}
in terms of the local basis, directly measuring the one magnon excitation. 
For EuCu3, we have the estimation for the interaction parameters, $J=10$~meV,
$D/J=0.3$, and the magnon peak position is $\Delta_{sw}=1.1J=88$~cm$^{-1}$, very
close to the measured value 72~cm$^{-1}$ in our Raman measurement of the
one-magnon peak.

~

\textbf{Data Availability}
All data supporting the findings of this study are available from the corresponding authors upon reasonable request.

~

\textbf{Acknowledgments:}
{This work was supported by the National Key Research and Development Program of
  China (2016YFA0301204), the program for Guangdong Introducing Innovative and
  Entrepreneurial Teams (No. 2017ZT07C062), by Shenzhen Key Laboratory of
  Advanced Quantum Functional Materials and Devices (No.
  ZDSYS20190902092905285), and by National Natural Science Foundation of China
  (Grant No.11774143, 12004377 and 11874350), the CAS Key Research Program of
  Frontier Sciences (ZDBS-LYSLH004) and China Postdoctoral Science Foundation
  (2019TQ0317). P.A. Lee acknowledges support by the US Department of Energy
  under grant number DE-FG02-03ER46076. 
}

~

 \textbf{Competing Interests:}
 The Authors declare no competing financial or non-financial interests.
 
   ~
   
   \textbf{Author Contributions:}
J.W.Mei conceived the project. P.H.Tan directed  the experimental work of Raman. Y.Fu,
L.Wang, L.Huang, W.Jiang and Z.Hao synthesized single crystals of samples.
M.Lin and P.H.Tan designed the Raman experiments.  M.Lin, P.H.Tan and J.Zhang
performed Raman measurements. Q.Liu and J.Dai performed the SHG measurements. Y.Fu,
L.Wang, L.Huang and C.Liu performed and analyzed magnetic susceptibility and
heat capacity measurements. H.Zhang, X.Shi and J.W.Mei performed
first-principles calculations. J.W.Mei, Y.Fu, M.Lin and P.H.Tan analyzed the Raman data. P.A.Lee, J.W.Mei and F.Ye worked on the theory, and
wrote the manuscript with contributions and comments from all authors. 

\end{document}


\title{ Supplementary Information: Dynamic fingerprint of fractionalized excitations in single-crystalline  
  \ce{Cu3Zn(OH)6FBr}}

\newcommand{\beginsupplement}{%
        \setcounter{table}{0}
        \renewcommand{\thetable}{S\arabic{table}}%
        \setcounter{figure}{0}
        \renewcommand{\thefigure}{S\arabic{figure}}%
        \setcounter{equation}{0}
        \renewcommand{\theequation}{S\arabic{equation}}%
     }

\author{Ying Fu}
\thanks{These two authors contributed equally.}
\affiliation{Shenzhen Institute for Quantum Science and Engineering, and Department of Physics, Southern University of Science and Technology, Shenzhen 518055, China}
\author{Miao-Ling Lin}
\thanks{These two authors contributed equally.}
\affiliation{State Key Laboratory of Superlattices and Microstructures, Institute of Semiconductors, Chinese Academy of Sciences, Beijing 100083, China}
\affiliation{Center of Materials Science and Optoelectronics Engineering \& CAS Center of Excellence in Topological Quantum Computation, University of Chinese Academy of Sciences, 100049 Beijing, China}
\author{Le Wang}
\author{Qiye Liu}
\author{Lianglong Huang}
\author{Wenrui Jiang}
\author{Zhanyang Hao}

\author{Cai Liu}
\affiliation{Shenzhen Institute for Quantum Science and Engineering, and Department of Physics, Southern University of Science and Technology, Shenzhen 518055, China}
\author{Hu Zhang}
\author{Xingqiang Shi}
\affiliation{College of Physics Science and Technology, Hebei University,
Baoding 071002, China}
\author{Jun Zhang}

\affiliation{State Key Laboratory of Superlattices and Microstructures, Institute of Semiconductors, Chinese Academy of Sciences, Beijing 100083, China}
\affiliation{Center of Materials Science and Optoelectronics Engineering \& CAS Center of Excellence in Topological Quantum Computation, University of Chinese Academy of Sciences, 100049 Beijing, China}
\author{Junfeng Dai}
\author{Dapeng Yu}
\affiliation{Shenzhen Institute for Quantum Science and Engineering, and
  Department of Physics, Southern University of Science and Technology, Shenzhen
  518055, China}
\author{Fei Ye}
\affiliation{Shenzhen Institute for Quantum Science and Engineering, and
  Department of Physics, Southern University of Science and Technology, Shenzhen
  518055, China}
 \affiliation{Shenzhen Key Laboratory of Advanced Quantum Functional Materials
   and Devices, Southern University of Science and Technology, Shenzhen 518055, China}

\author{Patrick A. Lee}
\affiliation{Department of Physics, Massachusetts Institute of Technology,
  Cambridge, Massachusetts 02139, USA}
\author{Ping-Heng Tan}
\email{phtan@semi.ac.cn}
\affiliation{State Key Laboratory of Superlattices and Microstructures, Institute of Semiconductors, Chinese Academy of Sciences, Beijing 100083, China}
\affiliation{Center of Materials Science and Optoelectronics Engineering \& CAS Center of Excellence in Topological Quantum Computation, University of Chinese Academy of Sciences, 100049 Beijing, China}
 \author{Jia-Wei Mei}
 \email{meijw@sustech.edu.cn}
 \affiliation{Shenzhen Institute for Quantum Science and Engineering, and
   Department of Physics, Southern University of Science and Technology,
   Shenzhen 518055, China}
 \affiliation{Shenzhen Key Laboratory of Advanced Quantum Functional Materials
   and Devices, Southern University of Science and Technology, Shenzhen 518055, China}

\date{\today}
\maketitle
\beginsupplement

\tableofcontents

\newpage

\section{Crystal photograph and thermodynamic characterization}
\begin{figure}[h]
  \centering
  \includegraphics[width=0.65\columnwidth]{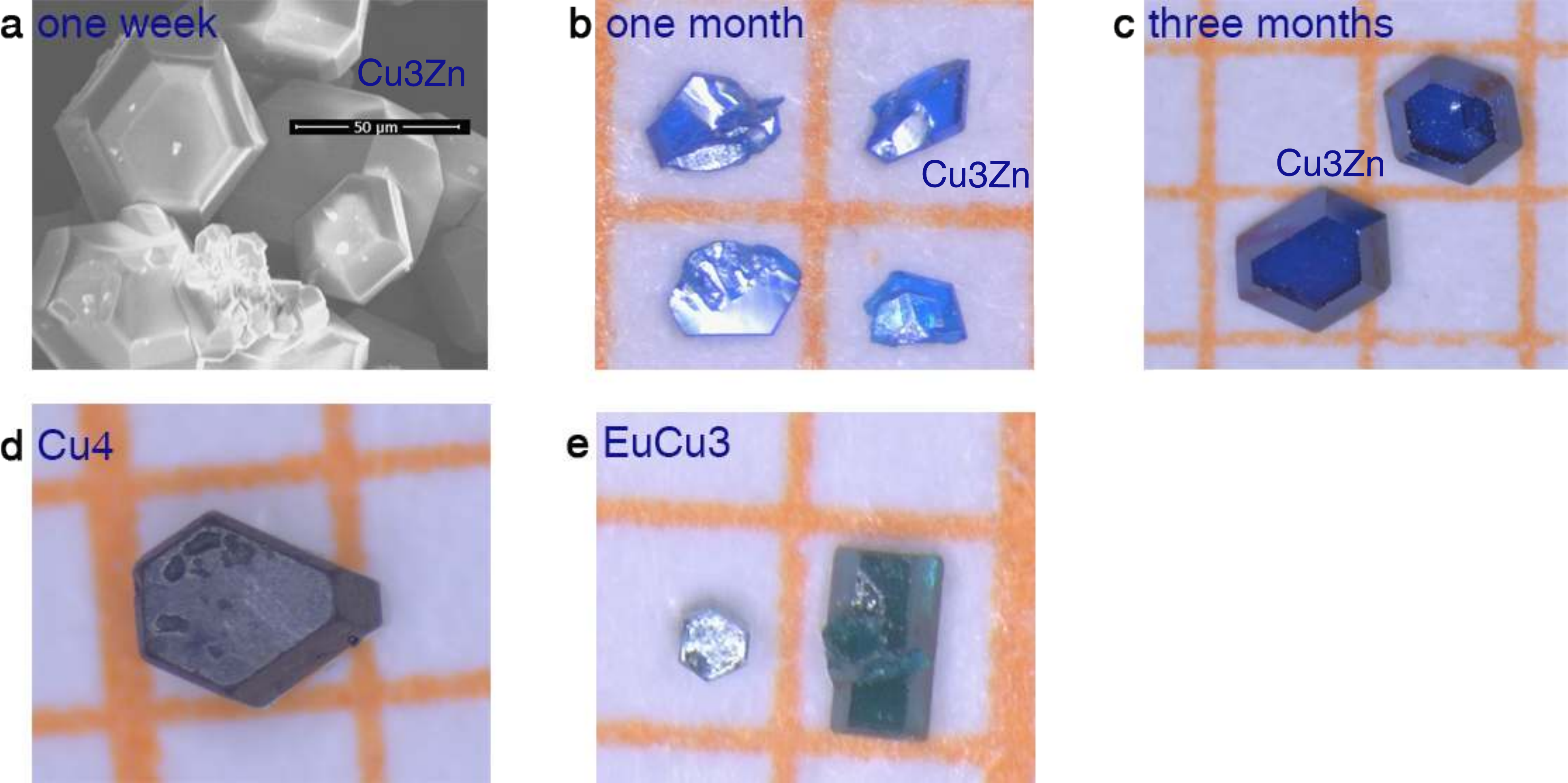}
  \caption{{\bf Photographs for single crystals of Cu3Zn, Cu4 and
      EuCu3.} Crystal sizes and morphologies of  \ce{Cu3Zn(OH)6FBr} (Cu3Zn) for different growth
    periods: ({\bf a}) one weak, ({\bf b}) one month, and ({\bf c}) three
    months. ({\bf d})
    \ce{Cu4(OH)6FBr} (Cu4); ({\bf e}) \ce{EuCu3(OH)6Cl3} (EuCu3). The yellow grid
  in ({\bf b}), ({\bf c}), ({\bf d}) and ({\bf e}) is 1$\times$1~mm$^2$.}
  \label{fig:figureS1}
\end{figure}

\begin{figure}[h]
  \centering
  \includegraphics[width=0.6\columnwidth]{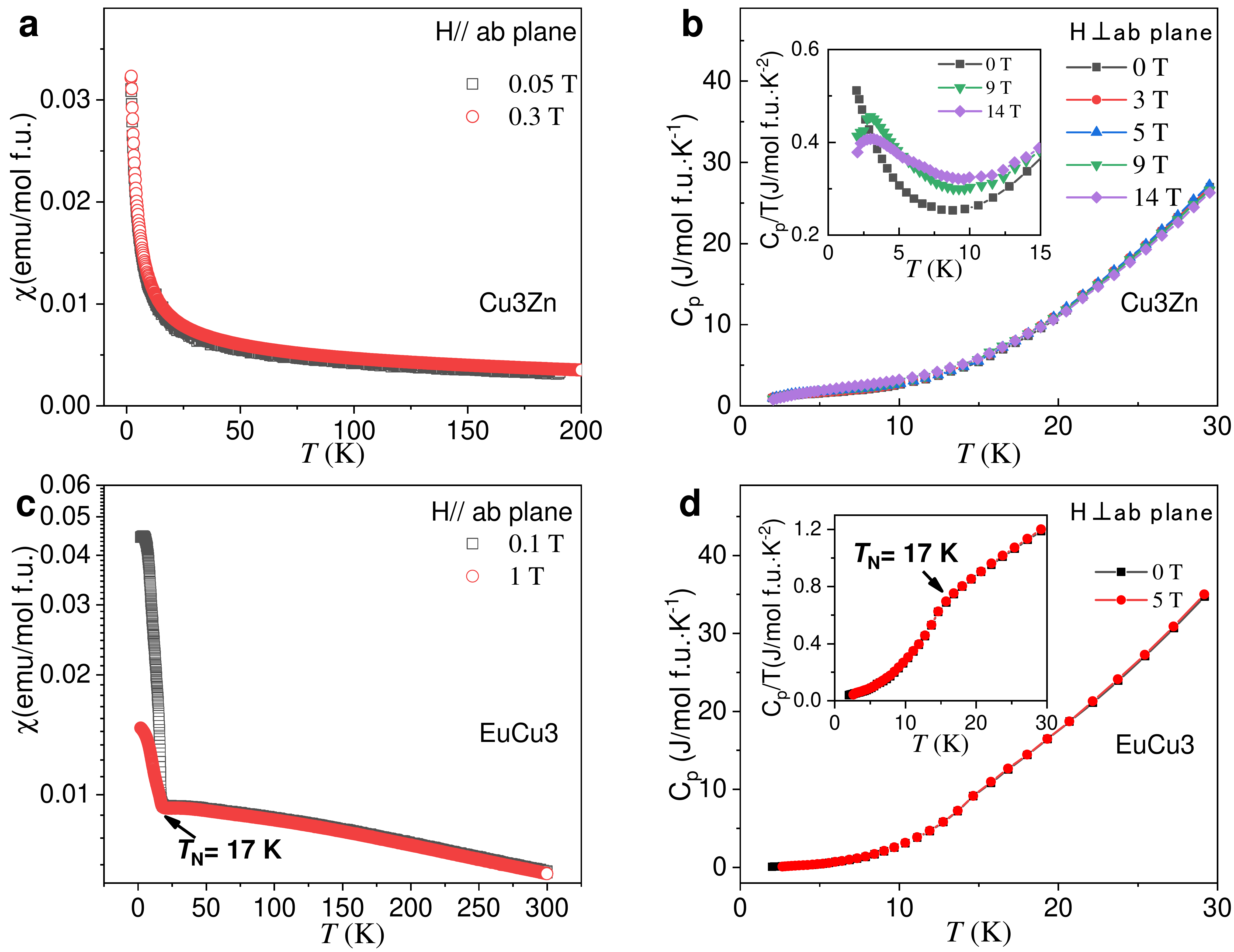}
  \caption{{\bf Thermodynamic properties of single crystals for Cu3Zn and EuCu3.}
    ({\bf a}) Temperature dependent magnetic susceptibilities ($\chi=M/H$) at
    0.05~T and 0.3~T fields.  ({\bf b})
    The temperature dependent specific heat C$_p$ at different magnetic fields
    in Cu3Zn. The thermodynamic properties of single crystals for Cu3Zn agree
    well with previous results on the powder samples.\cite{Feng2017,Wei2017} ({\bf c}) The magnetic susceptibilities show the ordering temperature
    $T_N=17$~K in EuCu$_3$. The Eu$^{3+ }$ with ground state of $^7F_0$
    contributes to the Van Vleck paramagnetism ({\bf d}) The temperature dependent heat capacities C$_p$  in EuCu3.}
  \label{fig:figureS2}
\end{figure}

\newpage
\section{Temperature evolution of the Raman spectra and phonon mode assignment in $\text{Cu3Zn}$}

\begin{figure}[h]
  \centering
  \includegraphics[width=0.7\columnwidth]{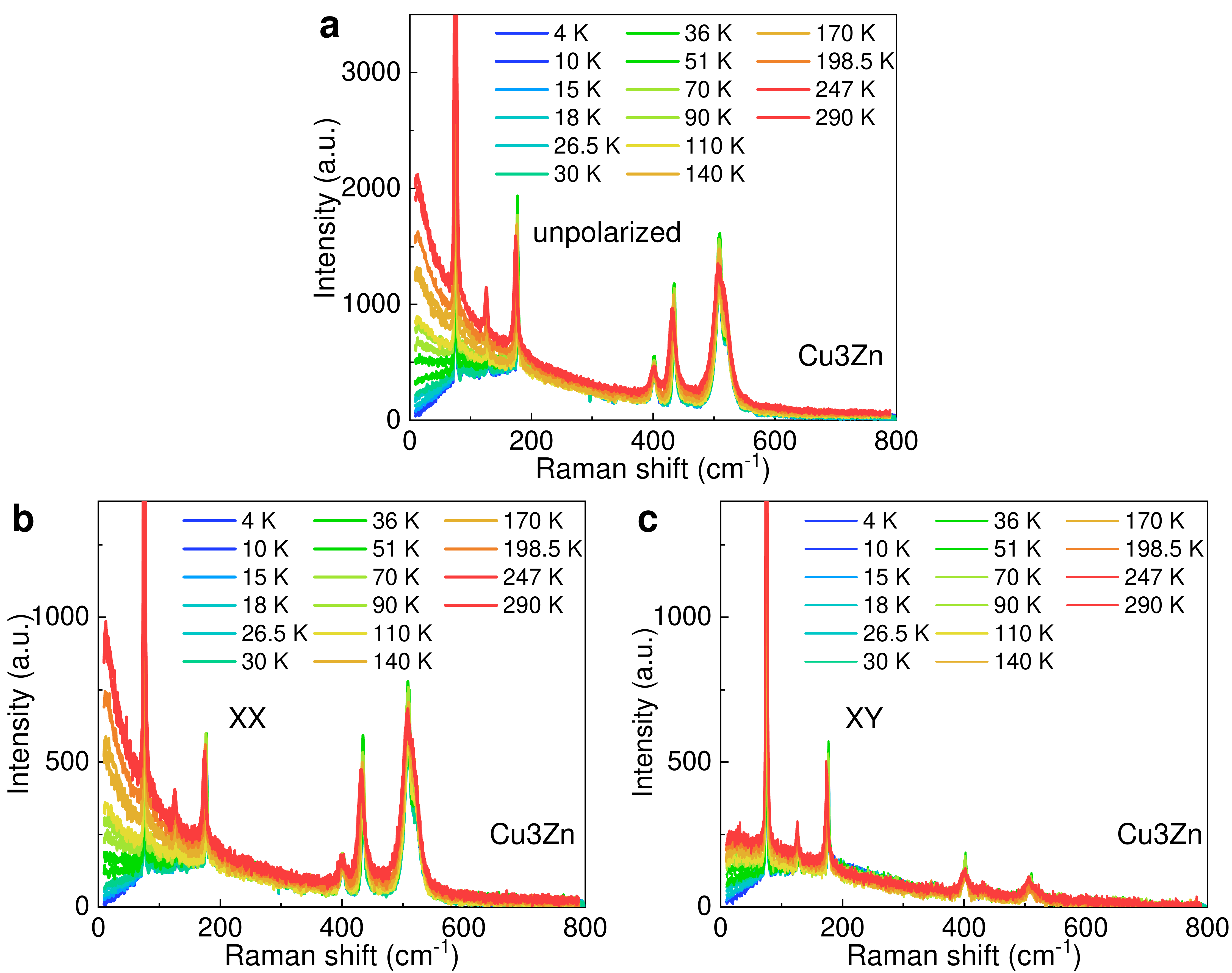}  
  \caption{{\bf Raman spectra in Cu3Zn at different temperatures.} ({\bf a}) Unpolarized
    Raman spectra in Cu3Zn. ({\bf b}) Raman spectra in the $XX$ configuration in
    Cu3Zn which contains the $A_{1g}$ and $E_{2g}$ channel. ({\bf c}) Raman spectra
    in the $XY$ configuration which contains the $E_{2g}$ channel in Cu3Zn.}
  \label{fig:figureS3}
\end{figure}

\begin{table}[h]
	\caption{Mode assignment for Cu3Zn.  Cu3Zn crystallizes the space group $P6_3/mmc$ (No.~194) and has Raman active
$A_{1g}$, $E_{1g}$, and $E_{2g}$ modes according to the point group
representation of $D_{6h}$ ($6/mmm$). $E_{1g}$ is not visible when the light
polarization lies in the kagome $ab$ plane, and we have Raman active phonon
modes $\Gamma_{\text{Raman}}=4A_{1g}+9E_{2g}$.}
	\begin{ruledtabular}
		\begin{tabular}{ccccc}
			 Frequency (Exp.) (cm$^{-1}$) &  Modes (Exp.) &Frequency (Cal.) (cm$^{-1}$)& Modes (Cal.) &Associated vibrating irons \\
			74.6  & $E_{2g}$&71.2  &$E_{2g}$ &Br$^-$ \\
			126.4 &$E_{2g}$ &124.11  &$E_{2g}$ &Zn$^{2+}$ \\
			172.2 &$E_{2g}$ &184.18 & $E_{2g}$& F$^-$\\
			355.5 & $E_{2g}$&345.37  & $E_{2g}$&O$^{2-}$ \\
			401.5 & $E_{2g}$&396.21  &$E_{2g}$ &O$^{2-}$ \\
			430.8 &$A_{1g}$ &426.06  & $A_{1g}$&O$^{2-}$ \\
			488.6 &$E_{2g}$, visible in 532~nm &493.47 & $E_{2g}$& O$^{2-}$ \\
			521.1  &$A_{1g}$ &508.84  & $A_{1g}$&O$^{2-}$ \\
			920.3  &$E_{2g}$&920.33  & $E_{2g}$&H$^+$ \\
			1016.7  &$A_{1g}$ &1082.26  &$A_{1g}$ & H$^+$\\
			1028.2  &$E_{2g}$, weak &1020.11 & $E_{2g}$& H$^+$\\
			3352.3 & ?&3333.19  &$E_{2g}$ &H$^+$ \\
			3467.5 &$A_{1g}$ & 3512.14 &$A_{1g}$ & H$^+$\\
		\end{tabular}
	\end{ruledtabular}
\end{table}

\newpage
\section{Raman spectra evolution from $\text{Cu4}$ to $\text{Cu3Zn}$ and Fano effect in $\text{Cu3Zn}$}
\begin{figure}[h]
\centering
\includegraphics[width=\columnwidth]{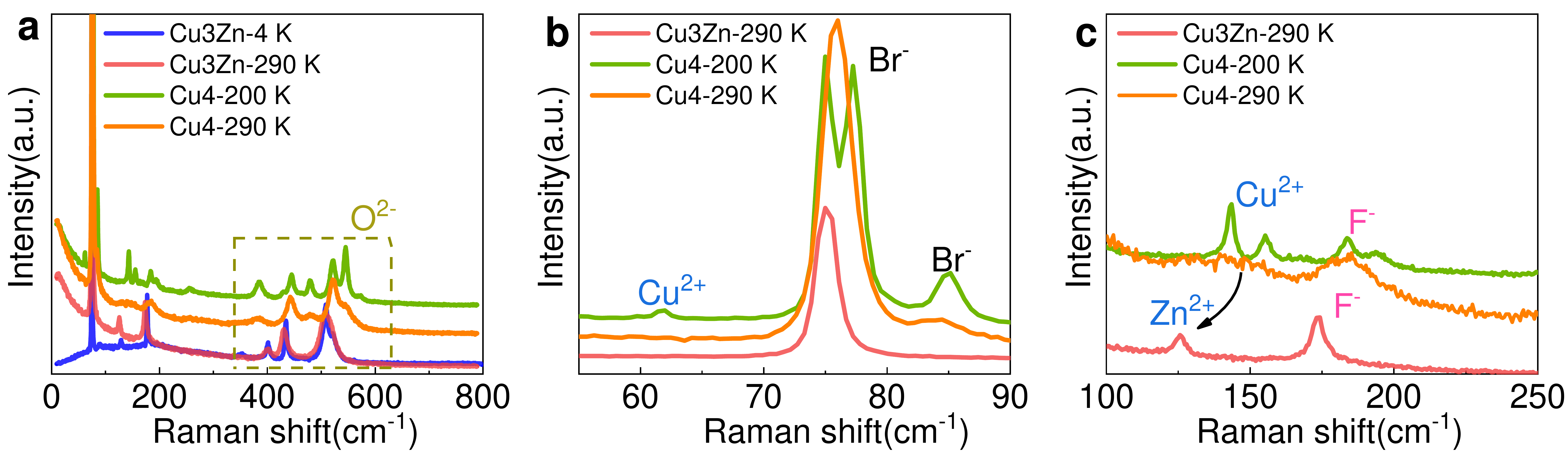}
\caption{\textbf{Raman spectral evolution from Cu4 to Cu3Zn} ({\bf a}) Unpolarized Raman spectra for Cu4 and
  Cu3Zn at selected temperatures.  
 Comparison for phonon modes between 40~cm$^{-1}$ and 90~cm$^{-1}$ in ({\bf b}),
 and between 100~cm$^{-1}$ and 250~cm$^{-1}$ in ({\bf c}) for 
Cu4 and Cu3Zn.   The Cu4 spectra in ({\bf a}), ({\bf b}) and ({\bf c})
    have been offset vertically for clarity. The phonon
evolution from Cu4 to Cu3Zn displays the difference by substituting Cu4
interlayer Cu$^{2+}$ site with Zn$^{2+}$ in Cu3Zn. The parent Barlowite Cu4  transforms to orthorhombic \textit{Pnma} below $T \approx 265$
K, characterized by changes in the relative
occupancies of the interlayer Cu$^{2+}$
site. Between 300~cm$^{-1}$ and 600~cm$^{-1}$, there are several phonon peaks
associated with O$^{2-}$ vibrations in Cu4 and Cu3Zn. Cu3Zn displays the Br$^{-}$ in-plane relative mode ($E_{2g}$) at
75~cm$^{-1}$, and has no
active Raman mode related to the kagome Cu$^{2+}$ vibrations since Cu$^{2+}$ is
the inversion center. The Br$^-$ phonon mode changes into two peaks in Cu4 due to the
superlattice folding in the orthorhombic Pnma phase at low temperature.
An additional Br$^-$ peak at 85~cm$^{-1}$ appears in Cu4, related to the Br
vibrations along the $c$-axis. 
The
kagome layers in Cu4 are distorted, signaled by a new phonon mode for the
kagome Cu$^{2+}$ vibration at 62~cm$^{-1}$.  Cu3Zn displays sharp $E_{2g}$ modes at 125~cm$^{-1}$
and 173~cm$^{-1}$ correspond to in-plane relative movements for Zn$^{2+}$ and
F$^{-}$, respectively. The corresponding modes (interlayer Cu$^{2+}$ and F$^{-}$) in Cu4
are broad at 290~K due to the randomly distributed interlayer Cu$^{2+}$ and split into
two peaks at 200~K. 
}
  \label{fig:figureS4}
\end{figure}

\begin{figure}[h]
  \centering
  \includegraphics[width=\columnwidth]{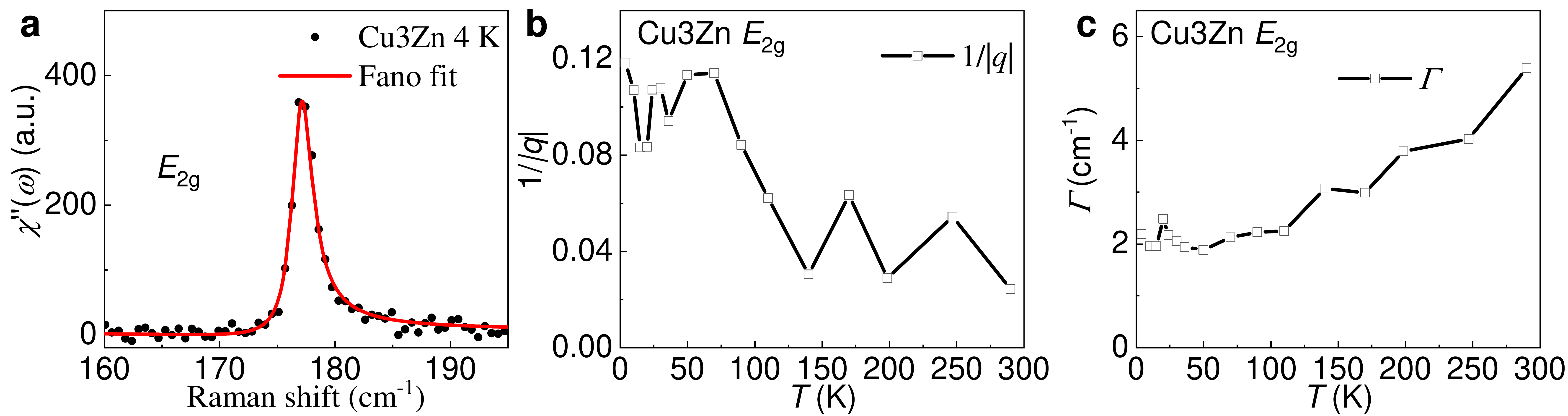}  
  \caption{{\bf  Fano lineshape of the $E_{2g}$ F$^-$  phonon peak
      at 173 cm$^{-1}$ in Cu3Zn.} ({\bf a}) Fano
    lineshape for the $E_{2g}$ in-plane F atomic movement phonon mode. 
    Temperature dependent Fano asymmetric parameter $1/|q|$ in ({\bf b}) and the width
    $\Gamma$ in ({\bf c}). The asymmetric Fano lineshape provides an additional probe of the magnetic degree of freedom.}
  \label{fig:figureS5}
\end{figure}

\newpage
\section{Angle-resolved  light polarization dependent Raman response for
  $\text{Cu3Zn}$}
Two typical polarization configurations were utilized to measure the
angle-resolved polarized Raman spectra: i) a half-wave plate was put after the
polarizer in the incident path to vary the angles between the polarization of
incident laser and the analyzer with the fixed vertical polarization, which can
be denoted as the $X$-only configuration; ii) a polarizer
is allocated in the common path of the incident and scattered light to
simultaneously vary their polarization directions, while the polarizations of
incident laser and analyzer were parallel or perpendicular to each other. By
rotating the fast axis of the half-wave plate with an angle of $\theta/2$, the
polarization of incident and/or scattered light is rotated by $\theta$.  
\begin{figure}[h]
  \centering
  \includegraphics[width=0.7\columnwidth]{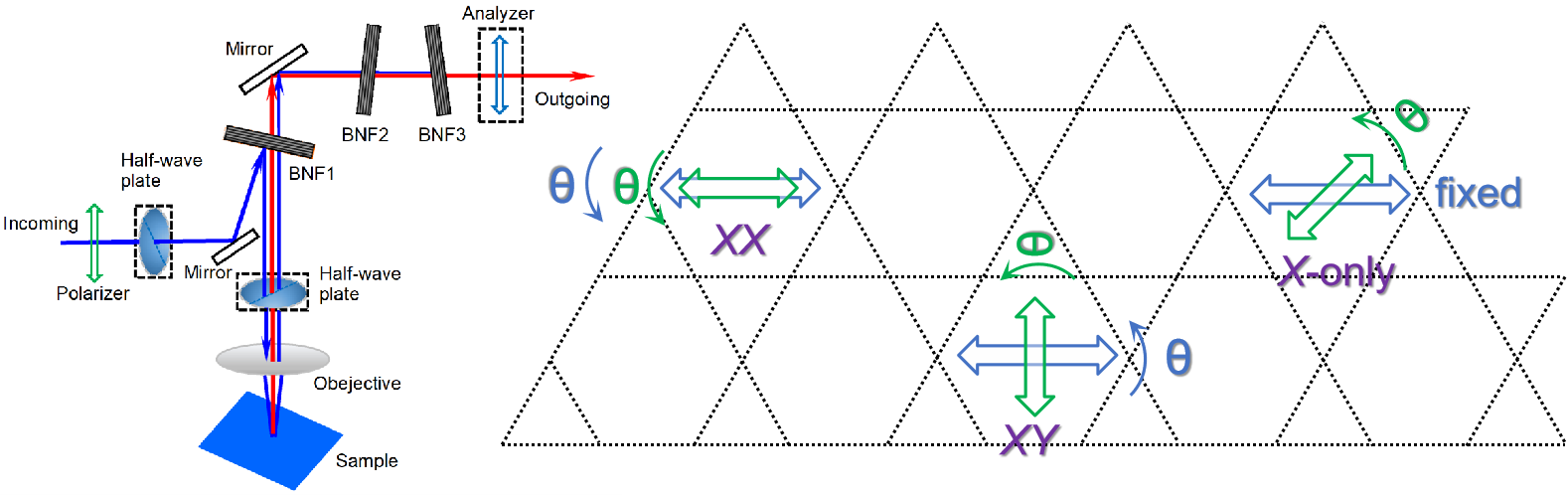}  
  \caption{{\bf Three polarization configurations in the angle
      dependent Raman response.} In the $XX$ ($XY$) configuration, the incoming and outgoing light
    polarizations are parallel (perpendicular) and we rotate both of them simultaneously. In the
  $X$-only configuration, the outgoing light polarization is fixed and we rotate
the incoming light polarization only.}
  \label{fig:figureS6}
\end{figure}

\begin{figure}[h]
  \centering
  \includegraphics[width=0.9\columnwidth]{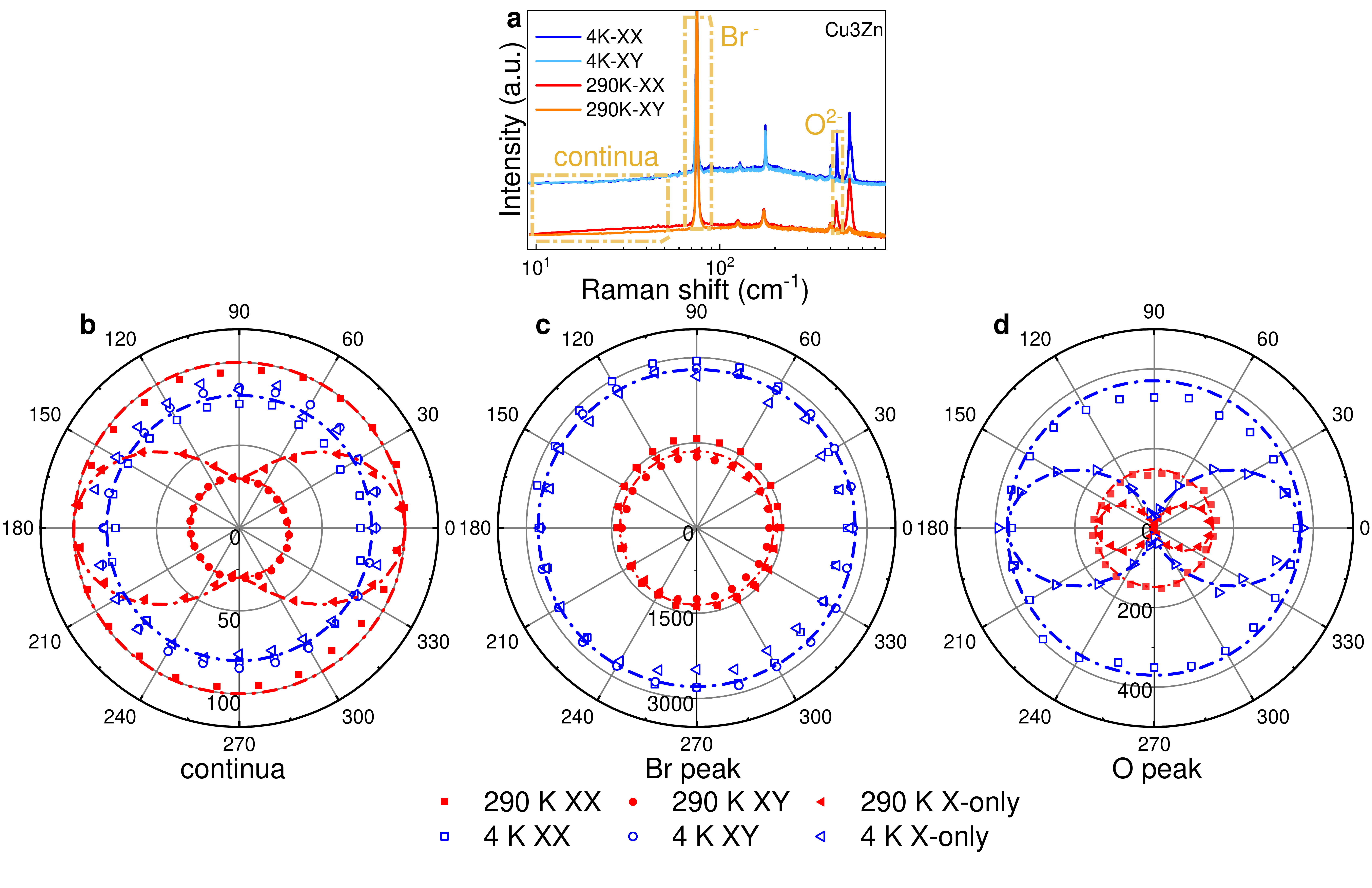}  
  \caption{{\bf Rotation symmetry of Raman dynamics for lattice vibrations and
    magnetic excitations in Cu3Zn. }
  ({\bf a}) We monitor three selected modes (both continua and phonon peaks). ({\bf b}) Angle dependence of the integrated Raman
  susceptibility
  $\chi_R=\frac{2}{\pi}\int_{10}^{60\text{~cm}^{-1}}\frac{\chi''(\omega)}{\omega}d\omega$.
  In $X$-only configuration, the continua at 290~K follows the $\cos^2(\theta)$
  for the $A_{1g}$ channel, while at other configurations, the continua remain constant.
 ({\bf c}) Angle dependence of the Br $E_{2g}$ phonon (75~cm$^{-1}$) scattering
  intensity. The lines are constant functions. ({\bf d}) Angle dependence of the O$^{2-}$ $A_{1g}$
  phonon (429~cm$^{-1}$) scattering intensity. The Raman intensity of O$^{2-}$
  $A_{1g}$ mode exhibits a $\cos^2(\theta)$ behavior in the $X$-only
  configuration at both room temperature and low temperature, and keeps constant
  in $XX$ and $XY$ configurations.
  }
  \label{fig:figureS7}
\end{figure}

\newpage
\section{Raman responses in $\text{EuCu3}$}
\begin{figure}[h]
  \centering
  \includegraphics[width=0.76\columnwidth]{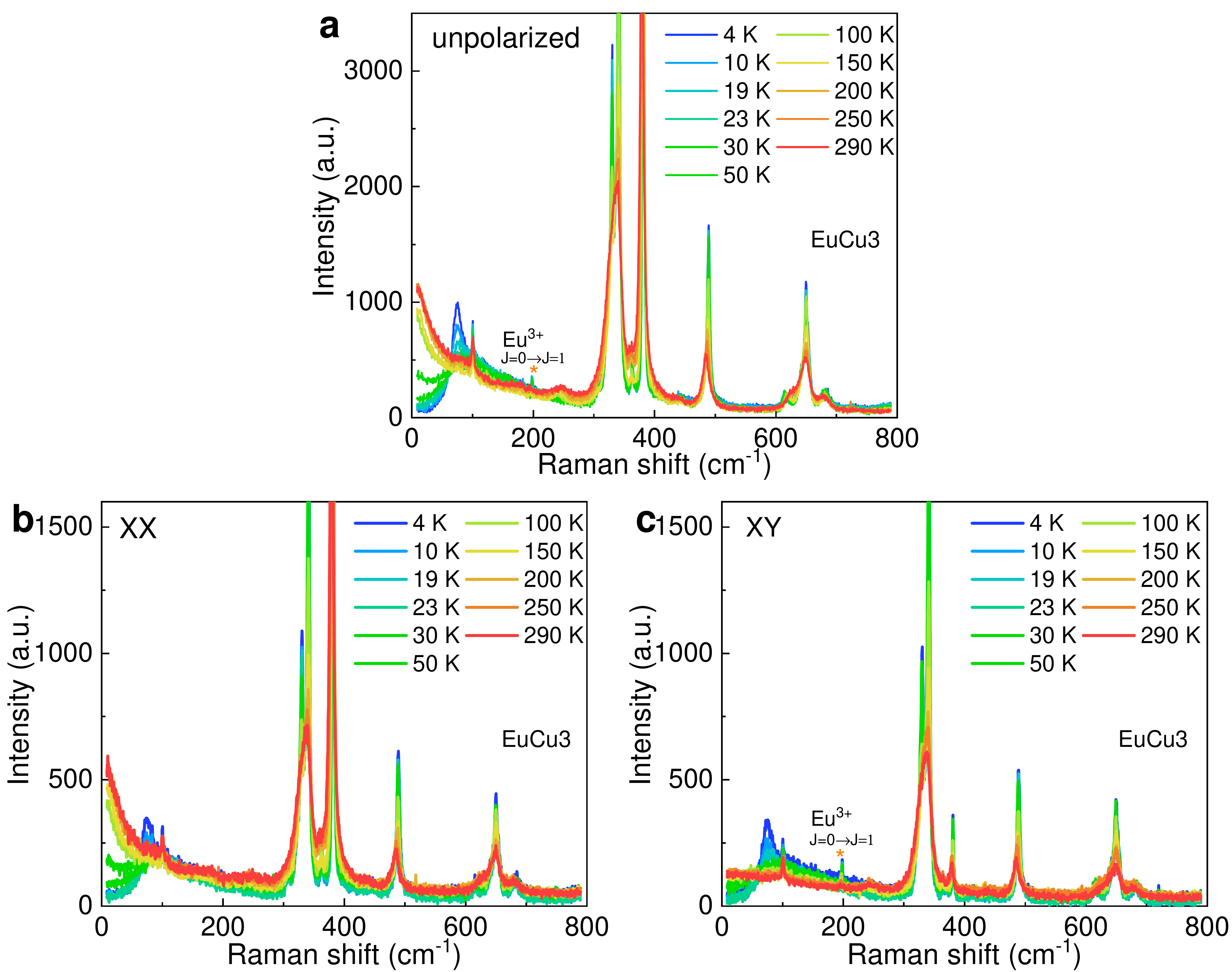}
  \caption{{\bf Raman spectra in EuCu3 at different temperatures.} ({\bf a}) Unpolarized
    Raman spectra in EuCu3. ({\bf b}) Raman spectra in the $XX$ configuration in
    EuCu3 which has the $A_{g}$ and $E_{g}$ channel. ({\bf c}) Raman spectra
    in the $XY$ configuration which contains the $E_{g}$ and $A_{2g}$ channel.
    For Eu$^{3+}$, we observe the $A_{2g}$ excitation of the $4f^6$
    configuration with the transition from $^7F_{J=0}$ to $^7F_{J=1}$.}
  \label{fig:figureS8}
\end{figure}

\begin{figure}[h]
  \centering
  \includegraphics[width=0.8\columnwidth]{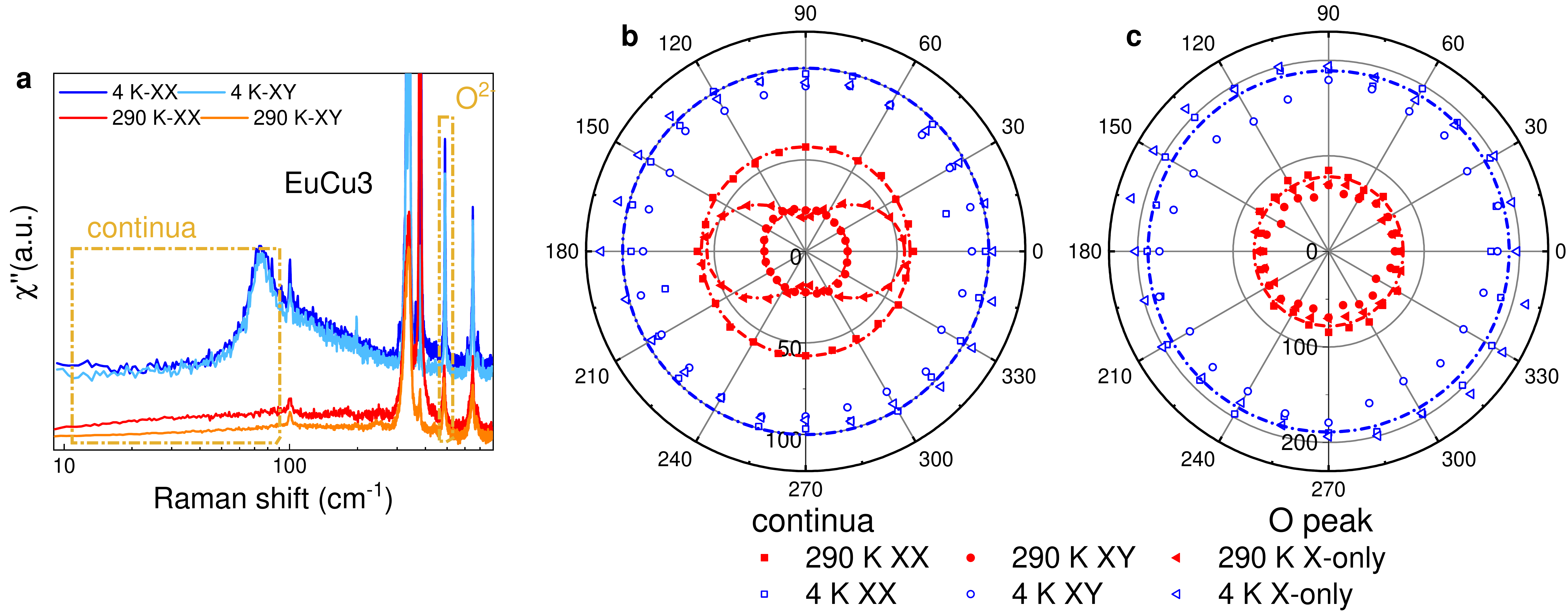}  
  \caption{{\bf Rotation symmetry of Raman dynamics for lattice vibrations and
    magnetic excitations in EuCu3. }
  We monitor the selected magnetic continuum at low frequency and the O$^{2-}$
  $E_g$ mode in ({\bf a}). ({\bf b})Angle
  dependence of the integrated Raman continuum from 9-80 cm$^{-1}$. The continua
  at 290~K follows $\cos^2(\theta)$ for the $A_{1g}$ channel, while others
  remain constant.  ({\bf c}) Angle dependence of the O$^{2-}$ $E_g$ phonon (487
  cm$^{-1}$) scattering intensity. Its Raman intensities are independent of $\theta$. }
  \label{fig:figureS9}
\end{figure}

\newpage
\section{Magnetic continuum in $\text{EuCu3}$ above $T_N$}

\begin{figure}[h]
  \centering
  \includegraphics[width=0.8\columnwidth]{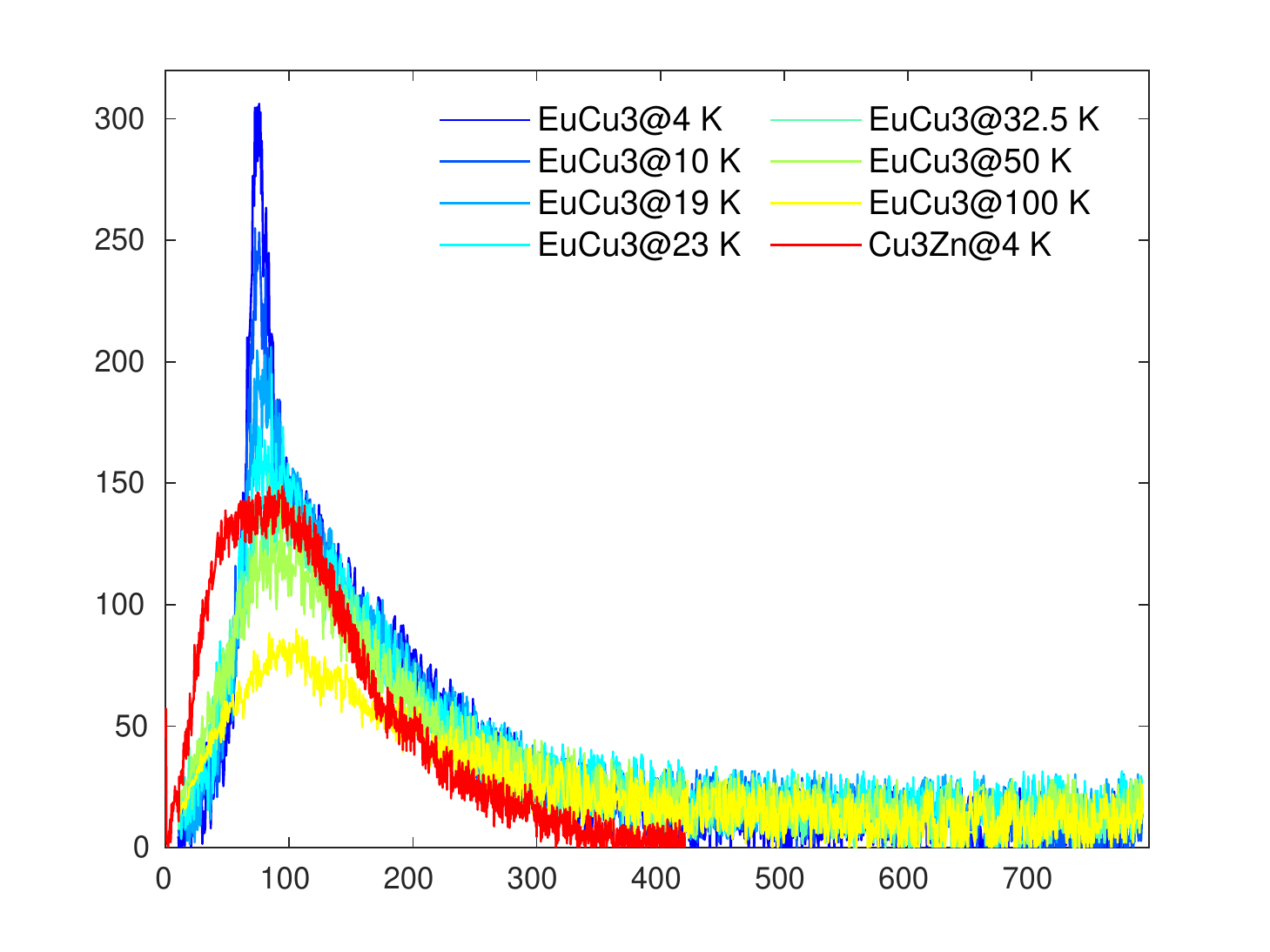}  
  \caption{{\bf Magnetic Raman susceptibility in the $XY$ configuration of EuCu3 above
      the N\'eel temperature.} We present the $XY$ magnetic Raman continuum in EuCu3
    below 100~K. Above  $T_N=17$~K, the Raman response has
    the substantial magnetic continuum below 50~K. For a comparison, we also
    plot the $XY$ magnetic Raman continuum in Cu3Zn at 4~K. The Raman shift
    frequency of  Cu3Zn is divided by 1.9, the ratio of the super-exchange strength
    of two compounds. We can see that above $T_N$, the profile of the Raman
    susceptibility in EuCu3 mimic that in Cu3Zn, suggesting the spinon
    contribution. There are less pronounced low-energy continuum excitations in
    EuCu3 than those in Cu3Zn, probably due to the large DM interaction which
    suppresses the low-energy quantum fluctuations. The maximum of the continuum
  excitations above $T_N$ in EuCu3 has the same energy scale as the magnon peak
  below $T_N$, which suggests that the magnon peak can be taken as the bound
  state of spinon-antispinon pair. }  
  \label{fig:figureS10}
\end{figure}
\newpage
\section{SHG of $\text{Cu3Zn}$}
\begin{figure}[h]
  \centering
  \includegraphics[width=0.45\columnwidth]{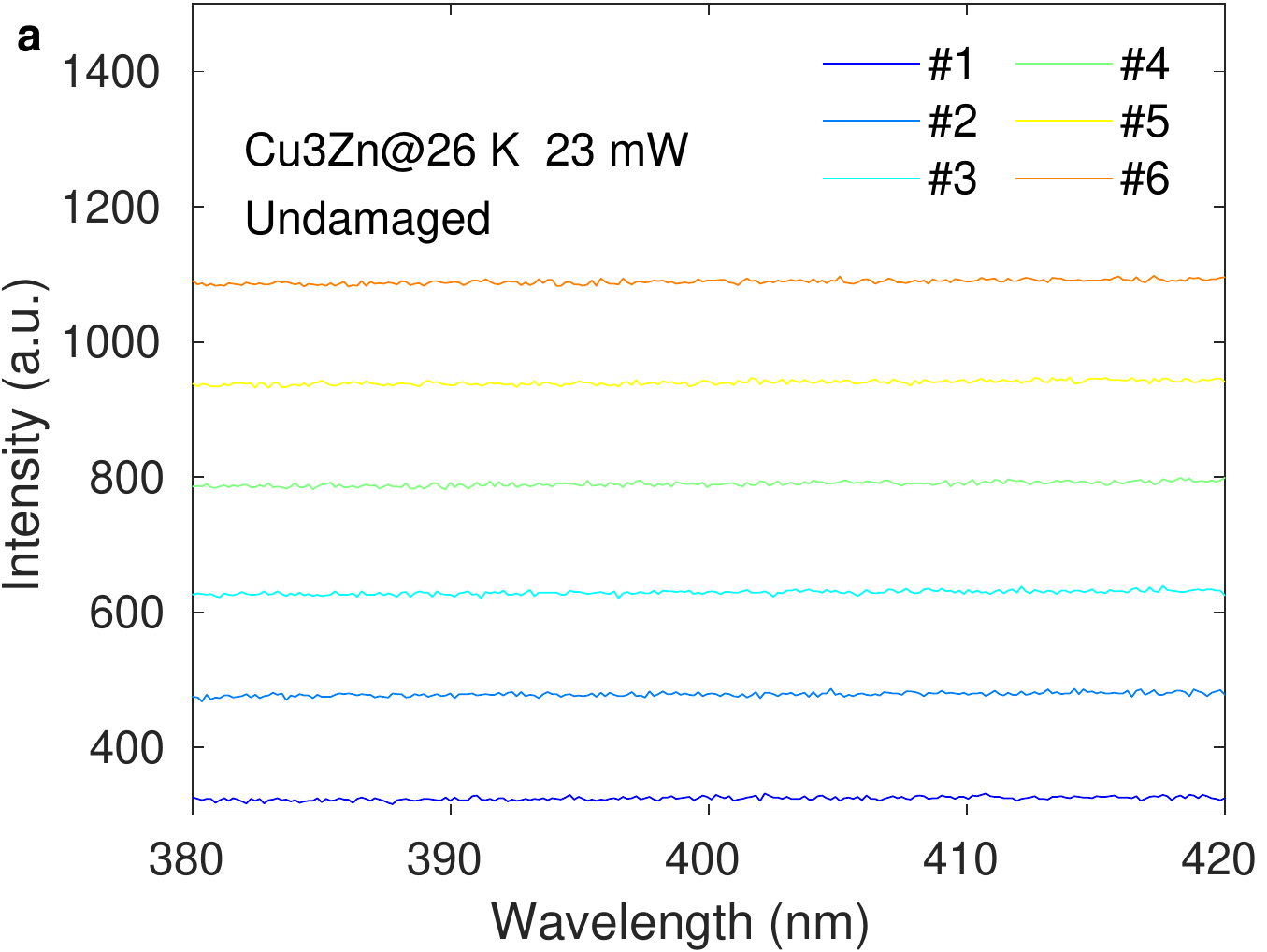} \includegraphics[width=0.45\columnwidth]{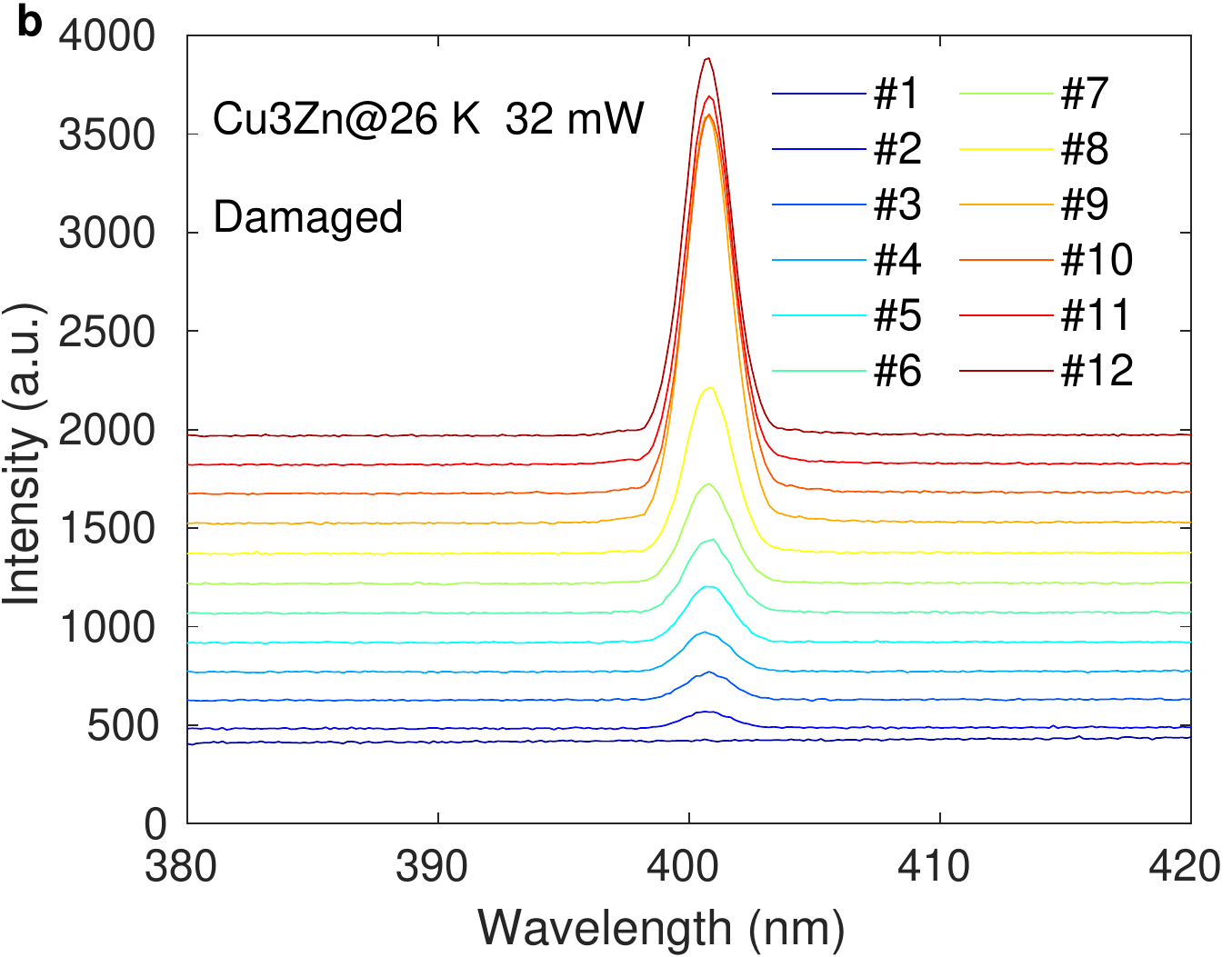}  
  \caption{{\bf SHG in Cu3Zn at 26 K with different laser powers.} ({\bf a}) SHG
    measurements in the same spot of sample taken every 5 seconds (from \#1 to
    \#6). At 23 mW, SHG signals in Cu3Zn sample are absent, implying that
    inversion symmetry remains preserved. ({\bf b}) A series of SHG measurements
    under the excitation power of 32 mW in the same point of the sample taken
    every 5 seconds (from \#1 to \#12). A remarkable SHG signal at 400 nm is
    detectable after a 10 second exposure, which dramatically enhances as the
    time increases. Due to the damage or degradation of Cu3Zn under high power
    excitation, the inversion symmetry breaking induces a strong SHG signals in
    sample. By comparison, we conclude that undamaged Cu3Zn single crystal
    presents spatial inversion symmetry at low temperature. The lines have been
    offset vertically for clarity. }  
  \label{fig:figureS11}
\end{figure}

\begin{figure}[h]
  \centering
  \includegraphics[width=0.45\columnwidth]{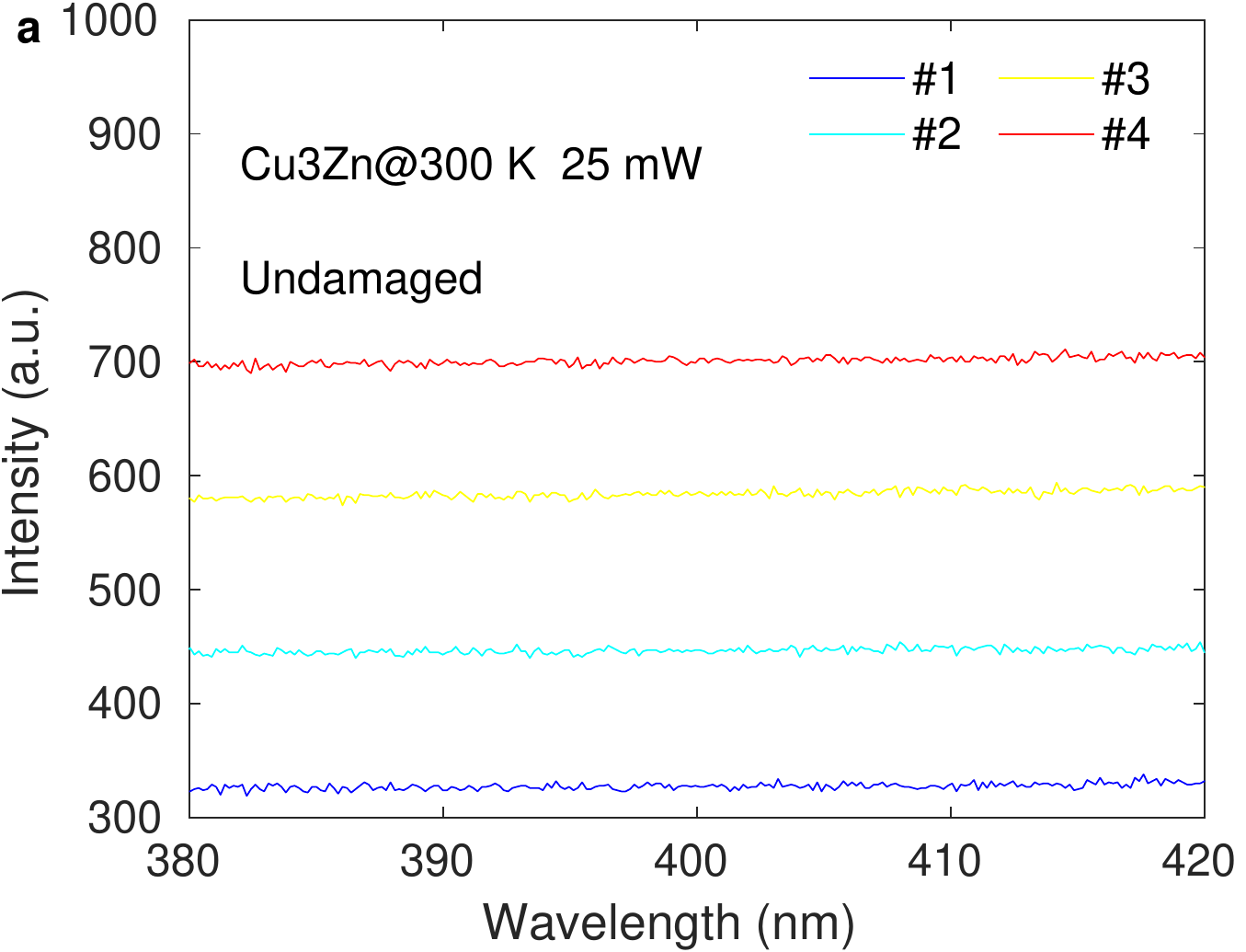} \includegraphics[width=0.45\columnwidth]{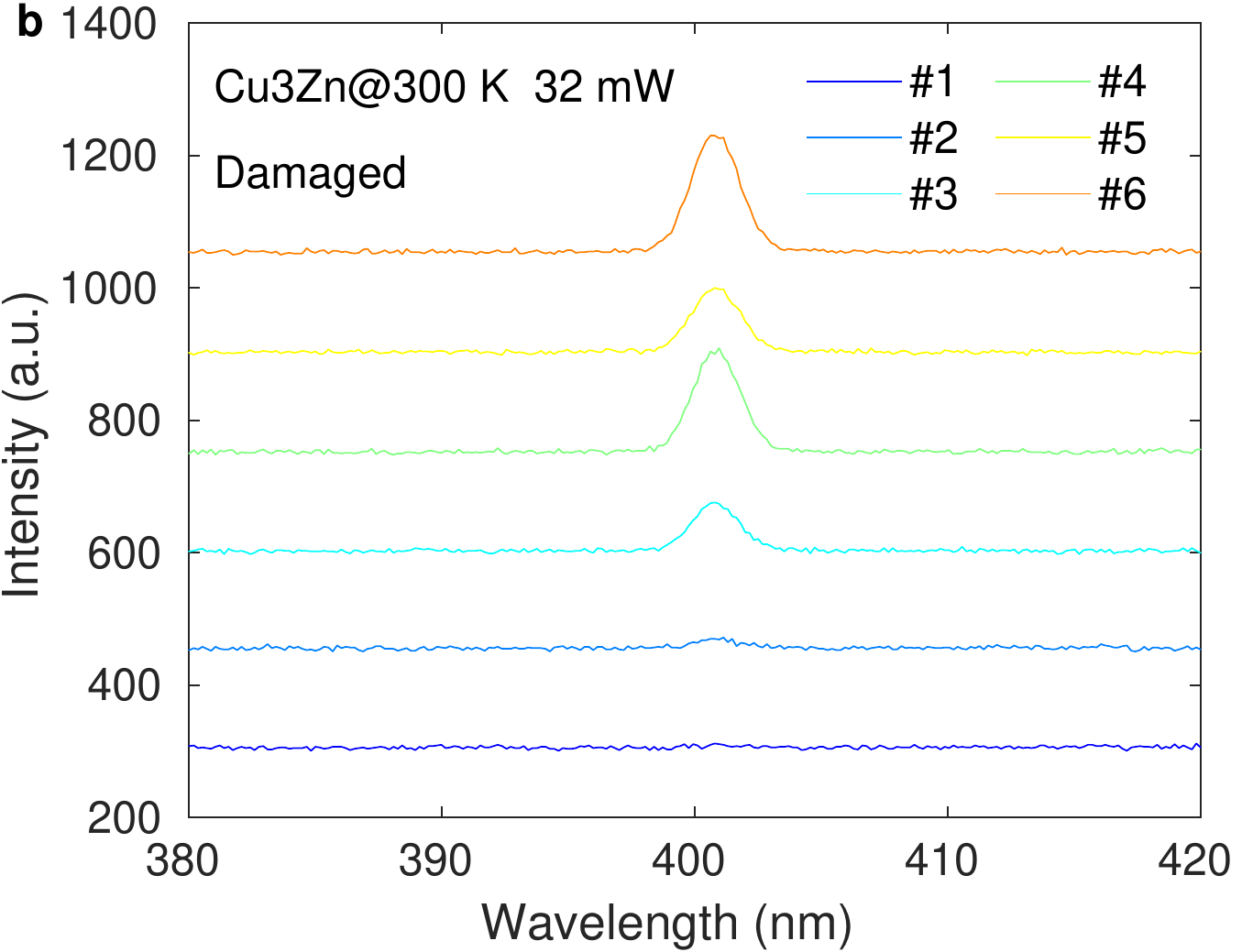}  
  \caption{{\bf SHG in Cu3Zn at 300 K with different laser powers.} ({\bf a})
    and ({\bf b})     represent the successive SHG measurements in the same point of sample taken
    every 5 seconds with excitation powers at 25 mW and 32 mW, respectively.
    There are no SHG signals at the excitation power of 25 mW, whereas strong
    SHG signals appear at the excitation power above 32 mW after a 10 second
    exposure. By comparison, damage or degradation in crystal structure under
    high power excitation induces a detectable SHG signal, implying that
    inversion symmetry presents in undamaged Cu3Zn at room temperature. The
    lines have been offset vertically for clarity.  }  
  \label{fig:figureS12}
\end{figure}
\newpage
\section{Estimation of exchange parameters in the kagome compounds}
 We implement the density functional theory (DFT)~\cite{Jones2015} to estimate the exchange
 parameters in the kagome compounds. We performed first-principles calculations
 with the Perdew–Burke–Ernzerhoff revised for solids (PBEsol) functional in
 generalized gradient approximation (GGA)~\cite{Perdew1996,Perdew2008} as
 implemented in the Vienna Ab Initio Simulation Package
 (VASP).\cite{Kresse1996,Gajdos2006,Hafner2008} An energy cutoff of 620~eV was
 used. We used $6\times6\times4$ Monkhorst-Pack grids~\cite{Monkhorst1976} for all
 calculations. All results were obtained with Cu 3$d$ valence electrons
 psudopotential within GGA+U (U$_{3d}$= 6~eV) scheme.\cite{Liechtenstein1995}

\begin{figure}[h]
  \centering
  \includegraphics[width=0.28\columnwidth]{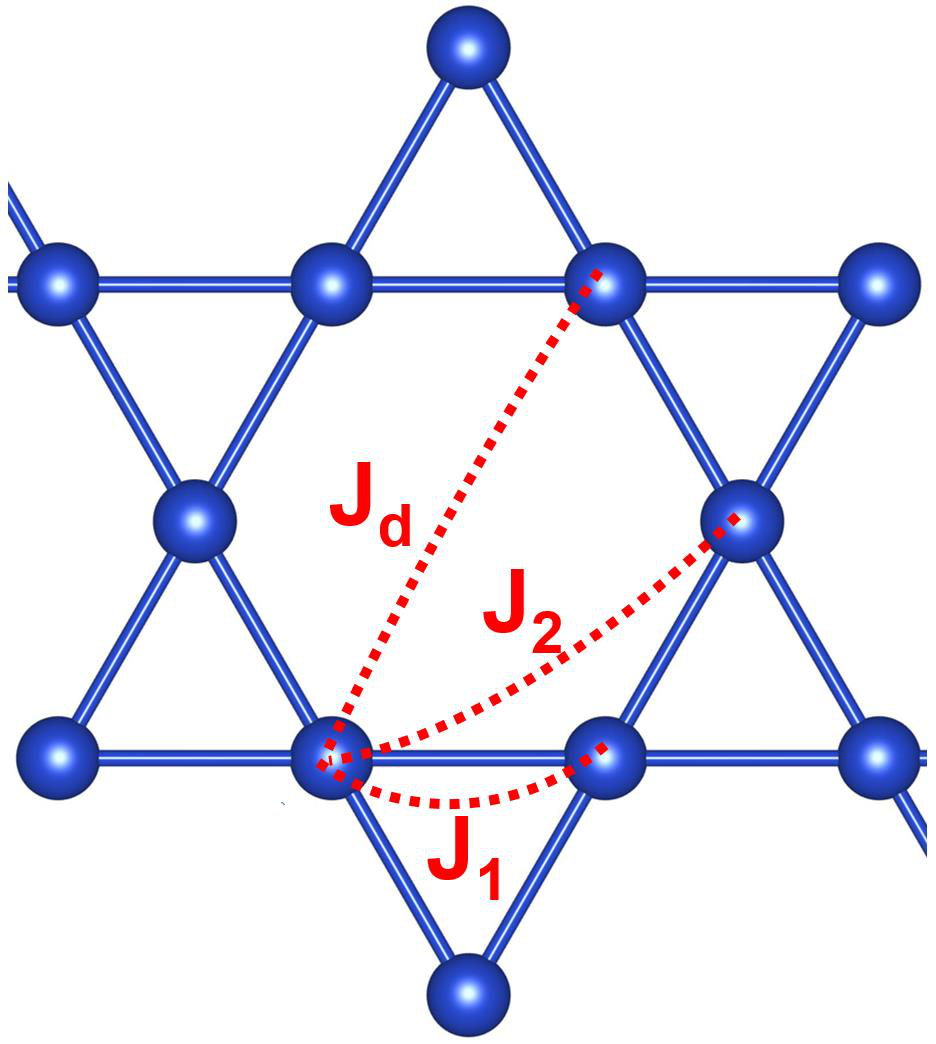}  
  \caption{{\bf Exchange interactions $J_1$, $J_2$, and $J_d$ in the Kagome lattice. }}  
  \label{fig:figureS13}
\end{figure}

We fix the lattice constants and relax the atomic positions with a coplanar magnetic structure
with negative vector spin chirality in the presence of spin-orbit couplings in our calculations. Exchange
interactions can be determined from total energies of virous different spin
configurations. To determine $J_1$, $J_2$, and $J_d$ (see
Fig.~\ref{fig:figureS13}), we used the
ferromagnetic state, antiferromagnetic state ($q=0$), cuboc2 state, and cuboc1
state for a $2\times2\times1$ supercell as discussed in Fig. 4 in Ref.~\cite{Messio2011}. 
Inter-layer couplings are determined by comparing the energies for different
stacking patterns of spin configurations. Note for the above interaction terms,
we turn off the spin-orbit coupling in our simulations. For the Dzyaloshinski-Moriya (DM)
interaction, we turn on spin-orbit couplings and compare the energies of $\mathbf{q}=0$ antiferromagnetic states
with positive and negative vector spin chiralities. The results are listed in
TAB.~\ref{tab:TabS2}. The nearest neighbor interactions in Cu3Zn and EuCu3 are
slightly larger than the experimental values.

\begin{table}[h]  
	\caption{Theoretical results of exchange interaction (in meV) and Cu-O-Cu
    bonding angle for various materials with the kagome structure. SG denotes
    the space group. $J_c$ denotes the inter-layer coupling. The references for the
  lattice constants are also listed.}\label{tab:TabS2}
	\begin{ruledtabular}
		\begin{tabular}{cccccccccc}
      Formula & SG & $J_1$ (meV)  & $J_2$ (meV) & $J_d$ (meV) & $J_c$ (meV) & DM (meV) & $\angle$Cu-O-Cu ($\circ$)& Reference\\
      \ce{Cu3Zn(OH)6FBr} & $P6_3/mmc$ & 24.13 & -0.01 & -0.65 & 1.442 & 1.12 & 117.47 & \cite{Feng2017}\\
      \ce{YCu3(OH)6Cl3} & $P\bar{3}m1$ & 10.21 & 0.12 & -0.09 & 0.040 & 3.45 & 118.60 & \cite{Zorko2019}\\
      \ce{EuCu3(OH)6Cl3} & $P\bar{3}m1$ & 13.02 & 0.16 & -0.08 & 0.003 & 3.83 & 120.33 & \cite{Puphal2018}\\
      \ce{SmCu3(OH)6Cl3} & $P\bar{3}m1$ & 13.55 & 0.14 & -0.08 & -0.004 & 5.91 & 120.36 & \cite{Sun2017}\\			
		\end{tabular}
	\end{ruledtabular}
\end{table}

\bibliography{../Refs}